\documentclass[prd,aps,showpacs,superscriptaddress,nofootinbib,amsmath,amssymb]{revtex4}

 \usepackage{graphicx}% Include figure files
\usepackage{dcolumn}% Align table columns on decimal point
\usepackage{bm}% bold math
\usepackage{lipsum}
\begin{document}
\title{Nuclear medium effects in $F_{2A}^{EM}(x,Q^2)$ and $F_{2A}^{Weak}(x,Q^2)$ structure functions}
%over a wide range of $x$ and $Q^2$}
\author{H. Haider}
\author{F. Zaidi\footnote{Corresponding author: zaidi.physics@gmail.com}}
\author{M. Sajjad Athar}
\author{S. K. Singh}
\affiliation{Department of Physics, Aligarh Muslim University, Aligarh - 202 002, India}
\author{I. \surname{Ruiz Simo}}
\affiliation{Departamento de F\'{\i}sica At\'omica, Molecular y Nuclear,
and Instituto de F\'{\i}sica Te\'orica y Computacional Carlos I,
Universidad de Granada, Granada 18071, Spain}
\begin{abstract}
Recent phenomenological analysis of experimental data on DIS processes induced by charged leptons and neutrinos/antineutrinos beams 
 on nuclear targets by CTEQ collaboration has confirmed the observation of CCFR and NuTeV collaborations, that weak 
 structure function $F_{2A}^{Weak} (x,Q^2)$ is different from 
 electromagnetic structure function $F_{2A}^{EM} (x,Q^2)$ in a nucleus like iron, specially in the 
region of low $x$ and $Q^2$. In view of this observation we have made a study of 
 nuclear medium effects on $F_{2A}^{Weak} (x,Q^2)$ and 
 $F_{2A}^{EM} (x,Q^2)$ for a wide range of $x$ and $Q^2$ using a microscopic nuclear model. We have considered Fermi motion,
binding energy, nucleon correlations, mesonic contributions from pion and rho mesons and shadowing effects to incorporate nuclear medium effects.
 The calculations are performed in 
a local density approximation using a
relativistic nucleon spectral function which includes nucleon correlations.
The numerical results in the case of iron nucleus are compared with the experimental data.
  \end{abstract}
\maketitle
\section{Introduction}
The Deep Inelastic Scattering(DIS) processes induced by charged leptons and neutrinos/antineutrinos on nucleons and nuclear 
targets are important tools to study the 
quark parton structure of free nucleons and the nucleons when they are bound in a nucleus.
The first observation of EMC effect in the DIS of muon~\cite{Aubert:1983xm} 
from iron target and using electron beam with various nuclear targets at SLAC~\cite{Bodek:1983qn}, discovered that
the quark parton distribution of free nucleons are considerably modified when they are bound in a nucleus. 
Subsequent DIS experiments performed on many nuclear targets with charged 
leptons~\cite{Arnold:1983mw, Aubert:1986yn, Whitlow:1990dr, Gomez:1993ri, Dasu:1993vk, Adams:1995sh, Arneodo:1997, Mamyan:2012th, Hen:2013oha, Malace:2014uea}
and neutrinos/antineutrinos~\cite{Berge:1989hr, Oltman:1992pq, Mousseau:2016snl, Tice:2014pgu, Morfin:2012kn, 
Varvell:1987qu, Seligman:1997mc, Fleming:2000bg, Onengut:2005kv, Tzanov:2005kr, Kuroda:2016hmr} have
confirmed the initial results of EMC effect. These experiments make measurements of DIS cross sections which are theoretically expressed in terms of
nucleon structure functions. 
The structure functions are determined by making Rosenbluth-like separations of the measured cross section. The 
electromagnetic DIS cross section induced by charged leptons are given in terms of two structure functions 
$F_i^{EM}(x,Q^2)$ (i=1,2), while the weak DIS cross section induced by neutrinos/antineutrinos are given in terms of three structure functions 
$F_i^{Weak}(x,Q^2)$ (i=1,2,3), where $x$ is the Bjorken scaling variable given by $x=\frac{Q^2}{2 M \nu}$; $Q^2$, $\nu$ and $M$ being the four momentum 
transfer square, the energy transfer($\nu=E_l-E_l^\prime$) to the target and nucleon mass, respectively~\cite{cooper_sarkar}. 
In the kinematic region of Bjorken 
scaling of large $\nu$ and $Q^2$, the structure functions $F_1(x,Q^2)$ and $F_2(x,Q^2)$ are related by the 
Callan-Gross relation~\cite{Callan:1969uq}. Therefore, in this kinematic region, 
the electromagnetic DIS cross sections are expressed 
in terms of only one structure function chosen to be $F_2^{EM}(x,Q^2)$ and the weak DIS cross sections are given in terms of two structure functions $F_2^{Weak}(x,Q^2)$ 
and $F_3^{Weak}(x,Q^2)$.
The DIS experiments induced by charged leptons have been done using nuclear targets in the entire region 
of mass number
from light nuclei to heavy nuclei, while the weak DIS experiments induced by neutrinos/antineutrinos have been done mainly 
on medium  and heavy 
nuclei~\cite{Varvell:1987qu, Seligman:1997mc, Fleming:2000bg, Onengut:2005kv, Tzanov:2005kr, Kuroda:2016hmr}. 
Therefore, it is very important to study the nuclear medium effects in the cross 
section and the structure functions determined from these DIS experiments in order to understand the
 structure functions of free nucleons and their modifications in nuclear medium.

Considerable theoretical and experimental efforts have been devoted to understand the nuclear medium effects on
the structure function $F_{2A}(x,Q^2)$ in electromagnetic and weak interactions as they play dominant role in 
determining the cross sections. A precise knowledge  of nuclear medium effects in $F_{2A}^{EM,Weak}(x,Q^2)$ can give information about the higher twist effects 
in the quark-parton distributions of both valence quarks and sea quarks in the free nucleon as well as their modifications in the
nuclear medium in contrast to $F_{3A}^{Weak}(x,Q^2)$ which is sensitive only to the valence quarks. Consequently 
a comparative study of 
$F_{2A}^{EM,Weak}(x,Q^2)$ and $F_{3A}^{Weak}(x,Q^2)$ gives very useful information about the properties of sea 
quarks and their 
modifications in nuclear medium.

 JLab is using high 
luminosity electron beams to make cross section measurements with several nuclear targets, which have  
resulted in obtaining wealth of new information 
on the structure functions $F_{1A}^{EM}(x,Q^2)$ and $F_{2A}^{EM}(x,Q^2)$ with good precision~\cite{Mamyan:2012th}. Furthermore, the future 
plan is to perform these measurements 
using high energy electron beam of 12~GeV in wide range of $x$ and $Q^2$. In the weak sector, several 
experiments are going on to study neutrino oscillation physics and 
some of them like MINERvA~\cite{Mousseau:2016snl, Tice:2014pgu} and DUNE~\cite{Strait:2016mof} are specially 
designed to precisely measure neutrino and antineutrino 
cross sections in the DIS region on some nuclear targets like 
carbon, oxygen, argon, iron and lead. The presently available data on electromagnetic nuclear structure functions from various 
experiments~\cite{Arnold:1983mw, Aubert:1986yn, Whitlow:1990dr, Gomez:1993ri, Dasu:1993vk, Adams:1995sh, Arneodo:1997, Mamyan:2012th, Hen:2013oha, Malace:2014uea}
have been phenomenologically analyzed to determine 
 the nuclear medium effects.  
 
 In the phenomenological analyses there are few approaches 
 for determining nuclear Parton Distribution Functions(PDFs). The general approach is that nuclear PDFs are obtained using the charged lepton-nucleus 
 scattering data and analyzing the ratio 
 of the structure functions e.g. $\frac{F_{2A}}{F_{2A^\prime}}$,  $\frac{F_{2A}}{F_{2D}}$, where $A, A^\prime$ represent any two nuclei and $D$ stands for the 
 deuteron, to determine nuclear correction factor. The nuclear correction factor is then multiplied with free nucleon PDFs to get nuclear PDFs.
 While determining the nuclear correction factor, the information regarding nuclear modification is also obtained from the Drell-Yan cross section ratio like  
 $\frac{\sigma_{pA}^{DY}}{\sigma_{pD}^{DY}}$, $\frac{\sigma_{pA}^{DY}}{\sigma_{p{A^\prime}}^{DY}}$, where p stands for proton beam. Furthermore, the information
 about the nuclear correction factor is also supplemented
 by high energy reaction data from the experiments at LHC, RHIC, etc.
 This approach has been used by Hirai et al.~\cite{Hirai:2007sx},
Eskola et al.~\cite{Eskola:2009uj}, Bodek and Yang~\cite{Bodek:2010km}, 
 de Florian and Sassot~\cite{deFlorian:2011fp}.
The same nuclear
correction factor is taken for the weak DIS processes. In a recent analysis de Florian et al.~\cite{deFlorian:2011fp} have also included 
 $\nu$-$A$ DIS data along with the charged lepton-nucleus 
 scattering data and Drell-Yan data. Their~\cite{deFlorian:2011fp} conclusion is that the same nuclear correction factor can describe the 
 nuclear medium effect in $l^{\pm}$-$A$ and $\nu$-$A$ DIS processes. 
 
 In the other approach nuclear PDFs are directly
 obtained by analyzing the experimental data i.e without using 
 nucleon PDFs or nuclear correction factor~\cite{Kovarik:2010uv}. This approach has been recently used by
 nCTEQ~\cite{Kovarik:2015cma, Kovarik:2012zz} group in getting $F_{2A}^{EM}(x,Q^2)$, $F_{2A}^{Weak}(x,Q^2)$ 
 and $F_{3A}^{Weak}(x,Q^2)$ by analyzing together the 
 charged lepton-$A$ DIS data and DY $p$-$A$ data sets, and separately analyzing
 $\nu(\bar \nu)-A$ DIS data sets. Their observation is that the nuclear medium effects on $F_{2A}^{EM}(x,Q^2)$ in
 electromagnetic
 interaction are different from $F_{2A}^{Weak}(x,Q^2)$ in weak interaction in the region of low $x$. Thus 
 in this region there is a disagreement 
 between the observation of these two studies~\cite{deFlorian:2011fp,Kovarik:2015cma}. The results of 
 CTEQ collaboration agree with the observation of CCFR 
 and NuTeV collaborations
 which had concluded that nuclear medium effects in $F_{2A}^{Weak}(x,Q^2)$ are different from $F_{2A}^{EM}(x,Q^2)$, 
 in the region 
 of $x~<~0.3$. This is the region in which nuclear medium effects as well as higher twist 
 effect even at the nucleon level may be important. The nuclear medium effects due to modifications of 
 sea quark distributions, 
 nuclear shadowing and mesonic contributions are important in this region.

 Theoretically, there have been very few calculations to study nuclear medium effects
 in the structure functions but no serious attempt has been made to make a comparison of 
  $F_{2A}^{EM}(x,Q^2)$ and $F_{2A}^{Weak}(x,Q^2)$ structure functions ~\cite{Donnachie:1993it, Qiu:2004qk, Kopeliovich:2004px, Kulagin:2004ie, Morfin:2012kn, Athar:2013gya}, therefore, 
 it is highly desirable to make a detailed study of nuclear medium effects on structure functions in this region.
 It has been pointed out that these nuclear structure functions are different in the region of 
 low $x$ due to nuclear shadowing effect being different
 in electromagnetic and weak processes.
 It is expected that the nuclear shadowing which is caused by the coherent scattering of hadrons produced in the 
 hadronization process of mediating vector bosons will be different in the case of electromagnetic and weak processes.
 This is because the electromagnetic and weak interactions take place
 through the interaction of photons and 
 $W^\pm/Z$ bosons, respectively, with the target hadrons and 
 the hadronization processes of photons and $W^\pm/Z$ bosons are different. Moreover, in the case of weak interaction, 
 the additional contribution of axial current which is not present in the case of electromagnetic interaction may 
 influence the 
behavior of $F_{2A}^{Weak}(x,Q^2)$ specially if pions also play a role in the hadronization process through PCAC.
Furthermore, in this region of low $x$, 
sea quarks and mesons also play important role which could be different in the case of electromagnetic and
weak processes.
 For example, the sea quark 
contribution, though very small is not same for $F_2^{EM}(x,Q^2)$ and $F_2^{Weak}(x,Q^2)$ even at 
the free nucleon
level and could evolve differently in a nuclear medium. A comparative study of mesonic contributions in the case of electromagnetic and weak structure functions  
should also be made in view of some recent work done for the DIS cross sections~\cite{Bodek:2010km, Haider:2011qs, prc85}.
% should also be made in view of some recent work in this direction~\cite{Bodek:2010km, Haider:2011qs, prc85}.

Therefore, a microscopic 
understanding of the difference between $F_{2A}^{EM}(x,Q^2)$ and $F_{2A}^{Weak}(x,Q^2)$ will be 
very instructive for studying the nuclear medium effects in DIS processes as emphasized recently 
at NuInt15~\cite{talknuint} workshop.

In this paper, we have studied nuclear medium effects in the structure functions using a microscopic approach based on field theoretical
 formalism~\cite{Marco:1995vb}. 
%  From the beginning a relativistic formalism has been used and Feynman diagrammatic
%  many body scheme has been followed, where all the inputs from the nucleons are contained in the nucleon 
%  propagator in the nuclear medium. 
 A relativistic nucleon spectral function has been used to describe the energy and momentum distribution of the
nucleons in nuclei~\cite{FernandezdeCordoba:1991wf}. This is obtained by using the Lehmann's representation for the relativistic nucleon
propagator and nuclear many body theory is used to calculate it for an interacting Fermi sea in
nuclear matter. A local density approximation is then applied to translate these results to
finite nuclei. Moreover, the contributions of the pion and rho meson clouds are also taken into account in a many
body field theoretical approach. This model has been earlier used by us to explain the 
nuclear medium effects 
in electromagnetic as well as weak interaction induced processes in the DIS 
region~\cite{sajjadnpa,Haider:2011qs,prc85,Haider:2012ic,Haider:2014iia,Haider:2015vea}.
 The present calculations are performed in a wide range of $x$ and $Q^2$ without using Callan-Gross relation between nuclear structure functions 
$F_{1A}(x, Q^2)$ and $F_{2A}(x, Q^2)$.  In Sect.~\ref{DIS_lepton_nucleon} some basic 
formalism for the inclusive charged lepton- and neutrino-nucleon/nucleus scattering has been introduced. 
In Sect.~\ref{sec:RE} the numerical results are presented and discussed for $F_{2A}^{EM}(x,Q^2)$ and $F_{2A}^{Weak}(x,Q^2)$. 
In Sect.~\ref{sec:Summary}, we 
conclude our findings.

\section{Nuclear Structure Functions} \label{DIS_lepton_nucleon}
\subsection{Interaction with a free nucleon target}
For the charged lepton induced deep inelastic scattering process 
($l(k) + N(p) \rightarrow l(k^\prime) + X(p^\prime);$ $l=~e^-,~\mu^-$), 
the differential scattering cross section is given by
\begin{equation}\label{eAf}
\frac{d^2 \sigma_N}{d\Omega_l dE_l^{\prime}} =~\frac{\alpha^2}{q^4} \; \frac{|\bf k'|}{|\bf k|} \;L_{\mu \nu} \; W_N^{\mu \nu},
\end{equation}
where $L_{\mu \nu}$ is the leptonic tensor and the hadronic tensor $W_N^{\mu \nu}$ is defined in terms of 
nucleon structure functions $W_{i N}$(i=1,3) as
\begin{eqnarray}\label{nuclearhtf}
L_{\mu \nu} &=& 2 [k_{\mu} k'_{\nu} +  k'_{\mu} k_{\nu} - k\cdot k^\prime g_{\mu \nu}] \nonumber\\
W_N^{\mu \nu} &=& 
\left( \frac{q^{\mu} q^{\nu}}{q^2} - g^{\mu \nu} \right) \;
W_{1N} + \left( p_N^{\mu} - \frac{p_N . q}{q^2} \; q^{\mu} \right)
\left( p_N^{\nu} - \frac{p_N . q}{q^2} \; q^{\nu} \right)
\frac{W_{2N}}{M_N^2}
\end{eqnarray}
with $M_N$ as the mass of nucleon. 

In terms of the Bjorken variable $x\left(=\frac{Q^2}{2M_N\nu}=\frac{Q^2}{2M_N (E_l - E_l^{\prime})}\right)$ and
$y\left(=\frac{\nu}{E_l}\right)$,
where $Q^2=-q^2$ and $\nu$ is the
energy transfer($=E_l-E_l^\prime$) to the nucleon in the Lab frame $ \left(\nu=\frac{p_N\cdot q}{M_N}=\frac{p^{ 0}_N q^0-p_{N}^z q^z}{M_N}\right)$, 
the differential cross section is given by
\begin{eqnarray}\label{diff_dxdy1}
\frac{d^2 \sigma_N}{d x d y}&=&
\frac{\pi \alpha^2 M_N^2 y}{2 E_l E_l^{\prime} sin^4\frac{\theta}{2}}
\left\{W_{2 N}(x, Q^2)cos^2\frac{\theta}{2}~+~2W_{1 N}(x, Q^2)sin^2\frac{\theta}{2}   \right\}\,.
\end{eqnarray}

Expressing in terms of dimensionless structure functions $F_{1N}(x,Q^2)=M_N W_{1 N}(\nu,Q^2)$ and $F_{2 N}(x,Q^2)=\nu W_{2 N}(\nu,Q^2)$, 
one may write
\begin{eqnarray}\label{diff_dxdy}
\frac{d^2 \sigma_N}{d x d y}&=&
\frac{8 M_N E_l \pi \alpha^2 }{Q^4}
\left\{xy^2 F_{1 N}(x, Q^2)
+ \left(1-y-\frac{xyM_N}{2 E_l}\right) F_{2 N}(x, Q^2)
\right\}\,.
\end{eqnarray}
  In the limit $Q^2 \rightarrow \infty$, $\nu \rightarrow \infty$, $x$ finite, the structure functions $F_{iN}(x, Q^2)(i=1,2)$  depend only 
on the  variable $x$  and satisfy the Callan-Gross relation~\cite{Callan:1969uq} given by $2xF_{1N}(x)=F_{2N}(x)$. 

The general expression of structure function $F_{2N}(x)$ for electromagnetic interaction is given by,
\begin{equation}
 F_2(x)=\sum_i e_i^2 x f_i(x),
\end{equation}
where $x$ is the fraction of parent's hadron momentum carried by a quark,
 $f_i(x)$ is the density distribution of quarks,
index $i$ runs over flavors of quarks and $e_i$ is the charge of the corresponding quark.
If we write the electromagnetic structure function for proton target then in a 4 flavor scheme it will be
\begin{eqnarray}
F_2^{ep}(x)&=&x\left[\frac{4}{9}(u(x) + \bar u(x))~+~\frac{1}{9} (d(x) + \bar d(x)) + \frac{1}{9}(s(x)+\bar s(x)) + \frac{4}{9}(c(x)+\bar c(x))\right]
\end{eqnarray}
Similarly, applying complete isospin symmetry between PDFs of neutrons and protons, the $F_2(x)$ structure function for neutron target can be written as
\begin{eqnarray}
F_2^{en}(x)&=&x\left[\frac{1}{9}(u(x) + \bar u(x))~+~\frac{4}{9} (d(x) + \bar d(x)) + \frac{1}{9}(s(x)+\bar s(x)) + \frac{4}{9}(c(x)+\bar c(x))\right]
\end{eqnarray}
For an isoscalar target,
\begin{eqnarray}
 F_2^{eN}(x)&=& \frac{F_2^{ep} + F_2^{en}}{2}= x\left[\frac{5}{18}(u(x) + \bar u(x))~+~\frac{5}{18}(d(x) + \bar d(x)) + \frac{1}{9} (s(x)+\bar s(x)) + \frac{4}{9}(c(x)+\bar c(x))  \right]
\end{eqnarray}

In the neutrinos/antineutrinos nucleon deep inelastic scattering process, neutrino interacts with 
a nucleon($N$), producing a lepton($l$) and jet of hadrons($X$) in the final state:
\begin{equation} 	\label{reaction}
\nu_l/\bar \nu_l(k) + N(p) \rightarrow l^{-}/l^{+}(k^\prime) + X(p^\prime),~l=e,~\mu
\end{equation}
In the above expression $k$ and $k^\prime$ are the four momenta of incoming neutrinos/antineutrinos and outgoing lepton respectively;
$p$ is the four momentum of the target nucleon and $p^\prime$ is the four momentum of hadronic state system $X$. This process is mediated by the 
exchange of virtual boson $W^\pm$ having four momentum $q$.

The double differential scattering cross section
evaluated for a nucleon target in its rest frame is expressed as:
\begin{equation} 	\label{ch2:dif_cross}
\frac{d^2 \sigma^{\nu/\bar\nu}_N}{d \Omega_l d E_l^{\prime}} 
= \frac{{G_F}^2}{(2\pi)^2} \; \frac{|{\bf k}^\prime|}{|{\bf k}|} \;
\left(\frac{m_W^2}{q^2-m_W^2}\right)^2
L_{\alpha \beta}^{\nu/ \bar\nu}
\; W^{\alpha \beta}_{N}\,,
\end{equation}
where ${G_F}$ is the Fermi coupling constant, $\Omega_l$ and $E_l^\prime$ refer to the outgoing lepton. 

The leptonic tensor for neutrinos/antineutrinos scattering $L^{\alpha \beta}_{\nu, \bar\nu}$ and the
 most general form of the hadronic tensor $W_{N}^{\alpha \beta}$ in terms of structure functions which are 
the functions of $\nu$ and $Q^2$ are given by
\begin{eqnarray}\label{ch2:had_ten_N}
L_{\alpha \beta}^{\nu/ \bar\nu}&=&k_{\alpha}k'_{\beta}+k_{\beta}k'_{\alpha}
-k.k^\prime g_{\alpha \beta} \pm i \epsilon_{\alpha \beta \rho \sigma} k^{\rho} 
k'^{\sigma}\,,\nonumber \\
W_{N}^{\alpha \beta} &=& 
\left( \frac{q^{\alpha} q^{\beta}}{q^2} - g^{\alpha \beta} \right) \;
W_{1N}^{\nu/\bar{\nu}}
+ \frac{1}{M_N^2}\left( p_N^{\alpha} - \frac{p_N . q}{q^2} \; q^{\alpha} \right)
\left( p_N^{\beta} - \frac{p_N . q}{q^2} \; q^{\beta} \right)
W_{2N}^{\nu/\bar{\nu}} \nonumber\\
&&-\frac{i}{2M_N^2} \epsilon^{\alpha \beta \rho \sigma} p_{N \rho} q_{\sigma}
W_{3N}^{\nu/\bar{\nu}} + \frac{1}{M_N^2} q^{\alpha} q^{\beta}
W_{4N}^{\nu/\bar{\nu}}  \nonumber\\
&&
+\frac{1}{M_N^2} (p_N^{\alpha} q^{\beta} + q^{\alpha} p_N^{\beta})
W_{5N}^{\nu/\bar{\nu}}
+ \frac{i}{M_N^2} (p_N^{\alpha} q^{\beta} - q^{\alpha} p_N^{\beta})
W_{6N}^{\nu/\bar{\nu}}\,,
\end{eqnarray}
where $W_{iN}^{\nu/\bar\nu}$ are the structure functions, which depend
on the scalars $q^2$ and $p_N.q$. In the limit $m_l \rightarrow 0$, the terms depending on $W_{4N}^{\nu/\bar\nu}$, 
$W_{5N}^{\nu/\bar\nu}$ and $W_{6N}^{\nu/\bar\nu}$ in Eq.~\ref{ch2:had_ten_N} do not contribute to the cross
section. 

The structure functions $W_{iN}^{\nu/\bar\nu} (x,Q^2)$(i=1,3) are redefined in terms of dimensionless 
structure functions $F_{iN}^{\nu/\bar\nu} (x)$ as: 
\begin{eqnarray}\label{ch2:relation}
M_N W_{1N}^{\nu/\bar{\nu}}(\nu, Q^2)=F_{1N}^{\nu/\bar{\nu}}(x),~~\nu W_{2N}^{\nu/\bar{\nu}}(\nu, Q^2)=F_{2N}^{\nu/\bar{\nu}}(x),~~ 
\nu W_{3N}^{\nu/\bar{\nu}}(\nu, Q^2)=F_{3N}^{\nu/\bar{\nu}}(x),
\end{eqnarray}

In the case of weak interaction, we have three structure functions $F_1(x),~F_2(x)$ and $F_3(x)$. The additional
structure function $F_3(x)$ is due to the vector-axial interference part of the weak interaction. $F_2^{Weak}(x)$ for 
neutrino and antineutrino induced processes on proton and neutron targets
are given by
\begin{eqnarray}
 F_2^{\nu p}&=& 2 x \left[d + s + \bar u + \bar c \right];~~
  F_2^{\bar\nu p}= 2 x \left[ u + c + \bar d + \bar s \right]\\
 F_2^{\nu n}&=& 2 x \left[ u + s + \bar d + \bar c \right];~~
  F_2^{\bar\nu n}= 2 x \left[ d + c +\bar u + \bar s \right]
\end{eqnarray} 
So, for an isoscalar target
\begin{eqnarray}
F_2^{\nu N}(x)=F_2^{\bar\nu N}(x)= x\left[ u(x) + \bar u(x) + d(x)  + \bar d(x) + s(x) + \bar s(x) + c(x) + \bar c(x)\right]
\end{eqnarray}

If we take the ratio of electromagnetic to weak structure functions with $Q^2$ dependence then
\begin{footnotesize}
\begin{eqnarray}\label{tzanov_eq}
\frac{F_2^{EM}(x,Q^2)}{F_2^{Weak}(x,Q^2)} &=& \frac{F_2^{eN}(x,Q^2)}{F_2^{\nu/\bar\nu N}(x,Q^2)}   = \it R(x,Q^2) 
 = \frac{5}{18}\left[1 - \frac{3}{5} \frac{s(x,Q^2) + \bar s(x,Q^2) - c(x,Q^2) -
\bar c(x,Q^2)}{\sum (q(x,Q^2) + \bar q(x,Q^2))} \right] 
\end{eqnarray}
\end{footnotesize}
If $s(x,Q^2) = \bar s(x,Q^2)$ and $c(x,Q^2) = \bar c(x,Q^2)$
\[
\frac{F_2^{EM}(x,Q^2)}{F_2^{Weak}(x,Q^2)}=\frac{F_2^{eN}(x,Q^2)}{F_2^{\nu/\bar\nu N}(x,Q^2)}=\frac{5}{18}\left[1 - \frac{6}{5} \frac{s(x,Q^2) - c(x,Q^2)}
{\sum (q(x,Q^2) + \bar q(x,Q^2))} \right] 
\]
and when one assumes $s(x,Q^2) = \bar s(x,Q^2) = c(x,Q^2) = \bar c(x,Q^2)$, then
\begin{equation}\label{equal}
 F_2^{EM}(x,Q^2) = \frac{5}{18} F_2^{Weak}(x,Q^2)
\end{equation}
Therefore, any deviation of $\it R(x,Q^2)$ from $\frac{5}{18}$ and/or any dependence on $x$, $Q^2$ will give information about the heavy quark 
distribution function in the nucleon.

\subsection{Interaction with a nuclear target}
For the charged lepton scattering taking place with a nucleon moving inside the nucleus, the expression of the  cross section is modified as
\begin{equation}\label{eAn}
\frac{d^2 \sigma_A}{d\Omega_l dE_l^{\prime}} =~\frac{\alpha^2}{q^4} \; \frac{|\bf k'|}{|\bf k|} \;L_{\mu \nu} \; W_A^{\mu \nu},
\end{equation}
where $W_A^{\mu \nu}$ is the nuclear hadronic tensor defined in terms of nuclear hadronic structure functions $W_{iA}$(i=1,2) as

\begin{figure}
 \includegraphics[height=5 cm, width=12 cm]{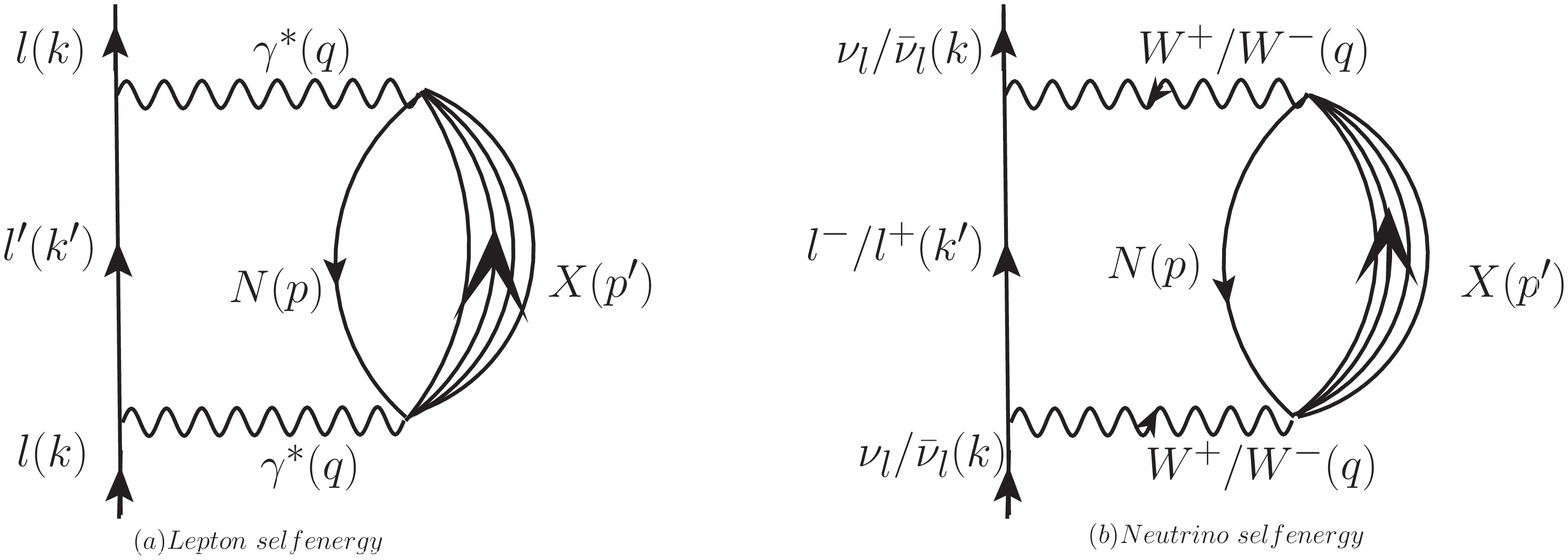}
 \caption{Diagrammatic representation of lepton self energy(left panel) and neutrino self energy(right panel).}
 \label{l_self}
\end{figure}
\begin{equation}\label{nuclearht}
W_A^{\mu \nu} = 
\left( \frac{q^{\mu} q^{\nu}}{q^2} - g^{\mu \nu} \right) \;
W_{1A} + \left( p_A^{\mu} - \frac{p_A . q}{q^2} \; q^{\mu} \right)
\left( p_A^{\nu} - \frac{p_A . q}{q^2} \; q^{\nu} \right)
\frac{W_{2A}}{M_A^2}
\end{equation}
with $M_A$ as the mass of the target nucleus.

In the following, we describe in brief the various nuclear medium effects, 
Fermi motion, binding, nucleon correlation,
isoscalarity correction and meson cloud contribution in obtaining the nuclear 
structure functions.

\subsubsection{Fermi motion, binding and nucleon correlation effects}

The cross section for an element of volume $dV$ in the nucleus is defined as~\cite{Marco:1995vb}:
\begin{equation}\label{defxsec}
d\sigma=\Gamma dt ds=\Gamma\frac{dt}{dl}ds dl=\Gamma \frac{1}{v}dV = \Gamma \frac {E_l}{\mid {\bf k} \mid}dV=\frac{-2m_l}{E_l({\bf k})} Im \Sigma (k)\frac{E_l({\bf k})}{\mid {\bf k} \mid}dV,
\end{equation}
where $\Gamma$ is the lepton width and $\Sigma(k)$ is the lepton self energy(shown in Fig.\ref{l_self}(a)). Thus 
to get $d\sigma$, we 
are required to evaluate
imaginary part of lepton self energy $Im \Sigma (k)$.

Lepton self energy $\Sigma (k)$ may be obtained by using Feynman rules as~\cite{Haider:2015vea}:
\begin{equation}\label{defn}
\Sigma (k) = i e^2 \; \int \frac{d^4 q}{(2 \pi)^4} \;
\frac{1}{q^4} \;
\frac{1}{2m_l} \;
L_{\mu \nu} \; \frac{1}{k'^2 - m_l^2 + i \epsilon} \; \Pi^{\mu \nu} (q),
\end{equation}
where $\Pi^{\mu \nu} (q)$ is the photon self energy.
Now the imaginary part of the lepton self energy can be readily obtained by applying Cutkosky rules
\begin{equation}\label{cut}
\begin{array}{lll}
\Sigma (k) & \rightarrow & 2 i \; I m \Sigma (k),~~~~~~~~~~~~~~~~~D (k') \rightarrow  2 i \theta (k'^0) \; I m D (k') \\
\Pi^{\mu \nu} (q) & \rightarrow & 2 i \theta (q^0) \; I m \Pi^{\mu \nu} (q), ~~~~~~~G (p) \rightarrow  2 i \theta (p^0) \; I m G (p) 
\end{array}
\end{equation}
which leads to
\begin{eqnarray}\label{self-lepton}
Im \Sigma (k) &=& e^2 \int \frac{d^3 q}{(2 \pi)^3} \;
\frac{1}{2E_l}\theta(q^0) \; Im(\Pi^{\mu\nu})  \frac{1}{q^4} \frac{1}{2m_l} \; L_{\mu \nu}\;
\end{eqnarray}

 It may be noted from Eq.~\ref{self-lepton} that $\Sigma (k)$ contains photon/W boson self energy $\Pi^{\mu\nu}$, which is 
written in terms of nucleon  propagator $G_l$ and meson  propagator $D_j$ and using Feynman rules this is given by
\begin{eqnarray}\label{photonse}
\Pi^{\mu \nu} (q)&=& e^2 \int \frac{d^4 p}{(2 \pi)^4} G (p) 
\sum_X \; \sum_{s_p, s_l} {\prod}_{\substack{i = 1}}^{^N} \int \frac{d^4 p'_i}{(2 \pi)^4} \; \prod_{_l} G_l (p'_l)\; \prod_{_j} \; D_j (p'_j)\nonumber \\  
&&  <X | J^{\mu} | H >  <X | J^{\nu} | H >^* (2 \pi)^4  \; \delta^4 (q + p - \sum^N_{i = 1} p'_i),\;\;\;
\end{eqnarray}
where $s_p$ is the spin of the nucleon, $s_l$ is the spin of the fermions in $X$, $<X | J^{\mu} | H >$ is the hadronic current for the initial state nucleon 
to the final state hadrons, index $l$ runs for fermions and index $j$ runs for bosons in the final hadron state $X$.

\begin{figure}
 \includegraphics[height=5 cm, width=10 cm]{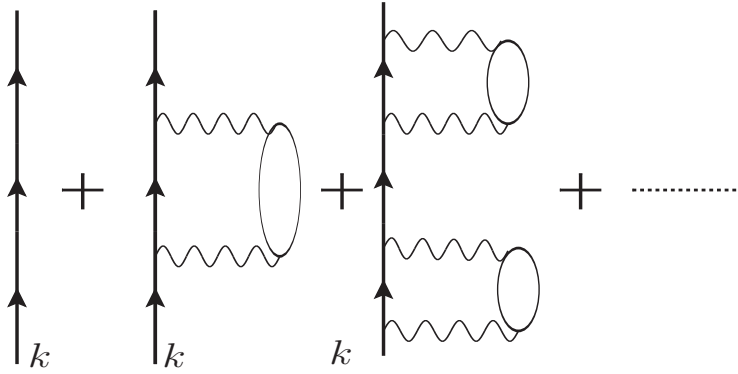}
 \caption{Diagrammatic representation of nucleon self energy in the nuclear medium.}
 \label{n_self}
\end{figure}

 The relativistic nucleon propagator $G(p)$ in a nuclear medium is obtained by starting with 
 the relativistic free nucleon Dirac propagator $G^{0}(p^{0},{{\bf p}})$ which is written in terms of the
contribution from the positive and negative energy components of the nucleon described by the Dirac spinors
$u({\bf p})$ and $v({\bf p})$~\cite{FernandezdeCordoba:1991wf,Marco:1995vb}. 
 Only the positive energy contributions are retained as the negative energy contributions are suppressed. 
 In the interacting 
Fermi sea, the relativistic nucleon propagator is then written using Dyson series expansion in terms of 
nucleon self energy $\Sigma^N(p^0,\bf{p})$ which is shown in Fig.\ref{n_self}. This perturbative 
expansion is summed in a ladder approximation as 
\begin{eqnarray}\label{gp1}
G(p)&=&\frac{M_N}{E({\bf p})}\frac{\sum_{r}u_{r}({\bf p})\bar u_{r}({\bf p})}{p^{0}-E({\bf p})}+\frac{M_N}{E({\bf p})}\frac{\sum_{r}u_{r}({\bf p})\bar
u_{r}({\bf p})}{p^{0}-E({\bf p})}\Sigma_{\alpha\beta}(p^{0},{\bf p})\frac{M_N}{E({\bf p})} \frac{\sum_{s}u_{s}({\bf p})\bar u_{s}({\bf p})}{p^{0}-E(P)}+..... \nonumber \\
&=&\frac{M_N}{E({\bf p})}\frac{\sum_{r} u_{r}({\bf p})\bar u_{r}({\bf p})}{p^{0}-E({\bf p})-\bar u_{r}({\bf p})\Sigma_{\alpha\beta}(p^{0},{\bf p})u_{r}({\bf p})\frac{M_N}{{E({\bf p})}}}
\end{eqnarray}
 where $\Sigma_{\alpha\beta}(p)$ is the nucleon self energy. The nucleon self energy $\Sigma_{\alpha\beta}(p)$ is 
 spin diagonal (=$\Sigma^N(p)\delta_{\alpha\beta}$), where $\alpha$ and $\beta$ are spinorial indices. 
 The nucleon self energy $\Sigma^N(p)$ is obtained following the techniques of standard many body theory and is taken from 
the works of spanish group~ \cite{FernandezdeCordoba:1991wf},\cite{Oset:1981mk} which uses the nucleon-nucleon scattering cross section 
 and the spin-isospin effective interaction with RPA correlation as inputs. In this approach
the real part of the self energy of nucleon is obtained by means of dispersion relations using the expressions for the imaginary part 
 which has been explicitly calculated. The Fock term which does not have imaginary part does not contribute either to $Im \Sigma$ or to $Re \Sigma$ 
 through the dispersion relation and its contribution to $\Sigma$ is explicitly calculated and added to $Re \Sigma$~\cite{FernandezdeCordoba:1991wf}.
 The model however misses some contributions from similar terms of Hartree type which are independent of nucleon momentum $p$. 
 This semi-phenomenological model of nucleon self energy is found to be in reasonable agreement with those obtained 
in sophisticated  many body calculations and has been successfully used in the past to study nuclear medium effects in many process
induced by photons, pions and leptons~\cite{Gil:1997bm}, \cite{Gil:1997jg}.
 
We use the expression for the nucleon self energy in nuclear matter i.e. $\Sigma^N(p^0,{\bf{p}})$ 
from Ref.~\cite{FernandezdeCordoba:1991wf}, and express
 \begin{eqnarray}\label{Gp}
G (p) =&& \frac{M_N}{E({\bf p})} 
\sum_r u_r ({\bf p}) \bar{u}_r({\bf p})
\left[\int^{\mu}_{- \infty} d \, \omega 
\frac{S_h (\omega, {\bf{p}})}{p^0 - \omega - i \eta}
+ \int^{\infty}_{\mu} d \, \omega 
\frac{S_p (\omega, {\bf{p}})}{p^0 - \omega + i \eta}\right]\,,
\end{eqnarray}
where $S_h (\omega, {\bf{p}})$ and $S_p (\omega, {\bf{p}})$ being the hole
and particle spectral functions respectively. Using Eqs.~ \ref{gp1}  and~\ref{Gp} above the expressions for 
the hole and particle spectral functions can be easily obtained which are given in Ref.\cite{FernandezdeCordoba:1991wf} and their properties are elaborated in some detail in Ref.\cite{Haider:2015vea}.

 The cross section is then obtained by using Eqs.~\ref{defn} and~\ref{self-lepton} : 
\begin{equation}\label{dsigma_3}
\frac {d\sigma_A}{d\Omega_l dE_l'}=-\frac{\alpha}{q^4}\frac{|\bf{k^\prime}|}{|\bf {k}|}\frac{1}{(2\pi)^2} L_{\mu\nu} \int  Im \Pi^{\mu\nu}d^{3}r, 
\end{equation}
where $\Pi^{\mu\nu}$ is obtained using Eqs.~\ref{photonse} and ~\ref{Gp}.

After performing some algebra, the expression of the nuclear hadronic tensor for an isospin symmetric nucleus in terms of 
 nucleonic hadronic tensor and spectral function, is obtained as~\cite{Haider:2015vea}
\begin{equation}	\label{conv_WA}
W^{\alpha \beta}_{A} = 4 \int \, d^3 r \, \int \frac{d^3 p}{(2 \pi)^3} \, 
\frac{M_N}{E ({\bf p})} \, \int^{\mu}_{- \infty} d p^0 S_h (p^0, {\bf p}, \rho(r))
W^{\alpha \beta}_{N} (p, q), \,
\end{equation}
where the factor of 4 is for spin-isospin of nucleon.

Accordingly the dimensionless nuclear 
structure functions $F_{iA}(x,Q^2)(i=1,2)$, are defined in terms of $W_{iA}(\nu,Q^2)(i=1,2)$ as
\begin{eqnarray}\label{relation1}
F_{2A}(x,Q^2)&=&\nu_A~W_{2A}(\nu,Q^2)~  {\rm where}\nonumber \\
\nu_A&=&\frac{p_{_A}\cdot q}{M_{_A}}=\frac{p^0_{_A} q^{0}}{M_{_A}}=q^{0},~~ p_{_A}^\mu=(M_{_A},\vec 0) 
\end{eqnarray}
For weak interaction, we follow the same procedure, and replace the lepton self energy diagram by neutrino
self energy diagram as shown in Fig.\ref{l_self}(b). We obtain the expression for electromagnetic 
and weak structure functions  $F_{_{2A}}^{EM,Weak}(x_A, Q^2)$ as 
\begin{eqnarray} \label{had_ten151}
F_{_{2A}}^{EM,Weak}(x_A, Q^2)  &=&  4 \int \, d^3 r \, \int \frac{d^3 p}{(2 \pi)^3} \, 
\frac{M_N}{E ({\bf p})} \, \int^{\mu}_{- \infty} d p^0 S_h (p^0, {\bf p}, \rho(r)) 
\times\left[\frac{Q^2}{(q^z)^2}\left( \frac{|{\bf p}|^2~-~(p^{z})^2}{2M_N^2}\right)\right. \nonumber \\
&& \left. +  \frac{(p^0~-~p^z~\gamma)^2}{M_N^2} \left(\frac{p^z~Q^2}{(p^0~-~p^z~\gamma) q^0 q^z}~+~1\right)^2\right]~\frac{M_N}{p^0~-~p^z~\gamma} ~F_{2N}^{EM,Weak}(x_N,Q^2),~~~       
\end{eqnarray}
where $x_N$ is the Bjorken variable for nucleon $\left( x_N=\frac{Q^2}{2 p_N \cdot q}=\frac{Q^2}{2(p^0_N q^0 - p_N^z q^z)}=\frac{Q^2}{2 q^0(p^0-\gamma p^z)} \right)$,
 which are moving  with $\gamma=\frac{q^z}{q^0}=
\left(1+\frac{4M^2x^2}{Q^2}\right)^{1/2}\,$. $x_A$ is the Bjorken variable for nucleus which is given by
$x_A=\frac{Q^2}{2 p_A \cdot q}=\frac{Q^2}{2 p^{0}_A q^0}= \frac{Q^2}{2 M_A q^0}=\frac{Q^2}{2 A M q^0}=\frac{x}{A}$. 
Three momentum of nucleus is zero(${\bf p_{A}}=0$) as in 
 the laboratory frame target nucleus is at rest and the nucleons are moving inside the nucleus. The nucleon structure
functions $F_{iN}(x,Q^2)$ (i=1-3), are expressed in terms of parton distribution functions(PDFs) which has been taken
from CTEQ6.6~\cite{Nadolsky:2008zw}.
 The evaluations are performed at the Next-to-Leading-Order(NLO) following the prescription 
 of Vermaseren et al.~\cite{Vermaseren:2005qc} and van Neerven and Vogt~\cite{vanNeerven:2000uj}. 
\subsubsection{Isoscalarity correction}
In many DIS experiments heavier nuclear targets are used where the neutron number (N) exceeds the proton 
number (Z), and the density of protons and neutrons are different. Historically, 
 the isoscalarity corrections in the case of electromagnetic processes 
induced by charged leptons are applied by multiplying the experimental results for the cross section by a factor $R_{ISO}^{A,EM}$, given by~\cite{Malace:2014uea} 
  \begin{eqnarray}\label{gg1}
 R_{ISO}^{A,EM}=\frac{(F_2^{lp}~+~F_2^{ln})/2}{[ZF_2^{lp}~+~(A-Z)F_2^{ln}]/A},\nonumber \\ 
 \end{eqnarray}
where $F_2^{lp}$ and $F_2^{ln}$ are the electromagnetic structure functions for proton and neutron, respectively.
 
 The resulting cross section per nucleon in the nuclear targets is then compared with the cross section for the isoscalar nucleon target which is obtained 
 from the deuteron data after applying deuteron corrections. This comparative study is used to determine the nuclear medium effects. 
 Phenomenologically the nuclear medium effect,
  thus obtained is parameterized by a function $f(x)$ (independent of $Q^2$), given by~\cite{Seligman:1997fe, Tzanov:2005kr}: 
  \begin{eqnarray}\label{ggfx}
 f(x)=\frac{F_{2A}^{EM}(x,Q^2)}{F_{2N}^{EM}(x,Q^2)}~
 = 1.0963~-~0.36427 x~-~0.27805 e^{-21.936 x}~+~2.7715~x^{14.417}
 \end{eqnarray}
 for isoscalar iron nuclear target and is used for many other nuclei in the medium mass range~\cite{Malace:2014uea}. In the above expression
 $F_{2A}^{EM}(x,Q^2)$ is structure function for an isoscalar nuclear target.
 
 This function $f(x)$ is also used to take into account nuclear medium effects for DIS by neutrinos/antineutrinos in nuclear targets, 
 after applying the following isoscalarity correction in the case of weak interaction processes
   \begin{eqnarray}\label{gg2}
  R_{ISO}^{A,Weak}=\frac{(F_2^{\nu/\bar\nu p}~+~F_2^{\nu/\bar\nu n})/2}{[ZF_2^{\nu/\bar\nu p}~+~(A-Z)F_2^{\nu/\bar\nu n}]/A},\nonumber \\ 
 \end{eqnarray}
 where $F_2^{\nu/\bar\nu p}$ and $F_2^{\nu/\bar\nu n}$ are the weak structure functions for proton and neutron, respectively.
  
In the present formalism, we take into account the nonisoscalarity of nuclear targets by considering separately
 hole spectral function for protons $S^{proton}_{h}$ and neutrons $S^{neutron}_{h}$ and normalizing these spectral functions to 
 respective proton and neutron number in a nuclear target.
To apply the present formalism for nonisoscalar nuclear target, we consider separate distributions of Fermi seas for protons and neutrons. 
 The nuclear hadronic tensor is modified to
\begin{eqnarray}\label{eq:convolution_hadronic_tensor_nonsymm_nuclear_matter}
W^{EM,Weak}_{\alpha\beta}&=&2\sum_{\tau=p,n} \int d^3r\int\frac{d^3p}{(2\pi)^3}\frac{M_N}{E(\mathbf{p})}\int^{\mu_\tau}_{-\infty}dp^0\; S^{\tau}_{h}
(p^0,\mathbf{p},p_{F,\tau}) W^{EM,Weak,\tau}_{\alpha\beta} 
\end{eqnarray}
where the factor $2$ in front of the integral accounts for the two degrees of freedom of the spin of the nucleons. 
  $p_{F,p}=(3\pi^2\rho_p)^{1/3}$ ($p_{F,n}=(3\pi^2\rho_n)^{1/3}$) is the Fermi momentum of proton (neutron). 
For the proton and neutron densities in heavy nuclear targets, we have used two-parameter Fermi density distribution 
and the density parameters are taken from Refs.~\cite{GarciaRecio, DeJager:1987qc}.
 
 For a nonisoscalar nuclear target the expressions for the structure functions $F_{_{2A}}^{EM,Weak}(x_A, Q^2)$ are obtained for the electromagnetic and weak interaction 
 processes as
\begin{eqnarray} \label{had_aten151}
F_{_{2A}}^{EM,Weak}(x_A, Q^2)  &=&  2\sum_{\tau=p,n} \int \, d^3 r \, \int \frac{d^3 p}{(2 \pi)^3} \, 
\frac{M_N}{E ({\bf p})} \, \int^{\mu_\tau}_{- \infty} d p^0 S_h^\tau (p^0, {\bf p}, \rho^\tau(r)) \times\left[\frac{Q^2}{(q^z)^2}\left( \frac{|{\bf p}|^2~-~(p^{z})^2}{2M_N^2}\right)\right. \nonumber \\
&& \left. +  \frac{(p^0~-~p^z~\gamma)^2}{M_N^2} \left(\frac{p^z~Q^2}{(p^0~-~p^z~\gamma) q^0 q^z}~+~1\right)^2\right]~\frac{M_N}{p^0~-~p^z~\gamma} ~F_{2N}^{EM,Weak,\tau}(x_N,Q^2),~~~       
\end{eqnarray}
while it is given by Eq.~\ref{had_ten151} for isoscalar nuclear targets. A comparison of Eq.~\ref{had_ten151} with Eq.~\ref{had_aten151} for
 $F_{_{2A}}^{EM, Weak}(x_A, Q^2)$, gives a quantitative estimate of isoscalarity effect in our model.
 \begin{figure}
 \includegraphics[height=12 cm,width=14 cm]{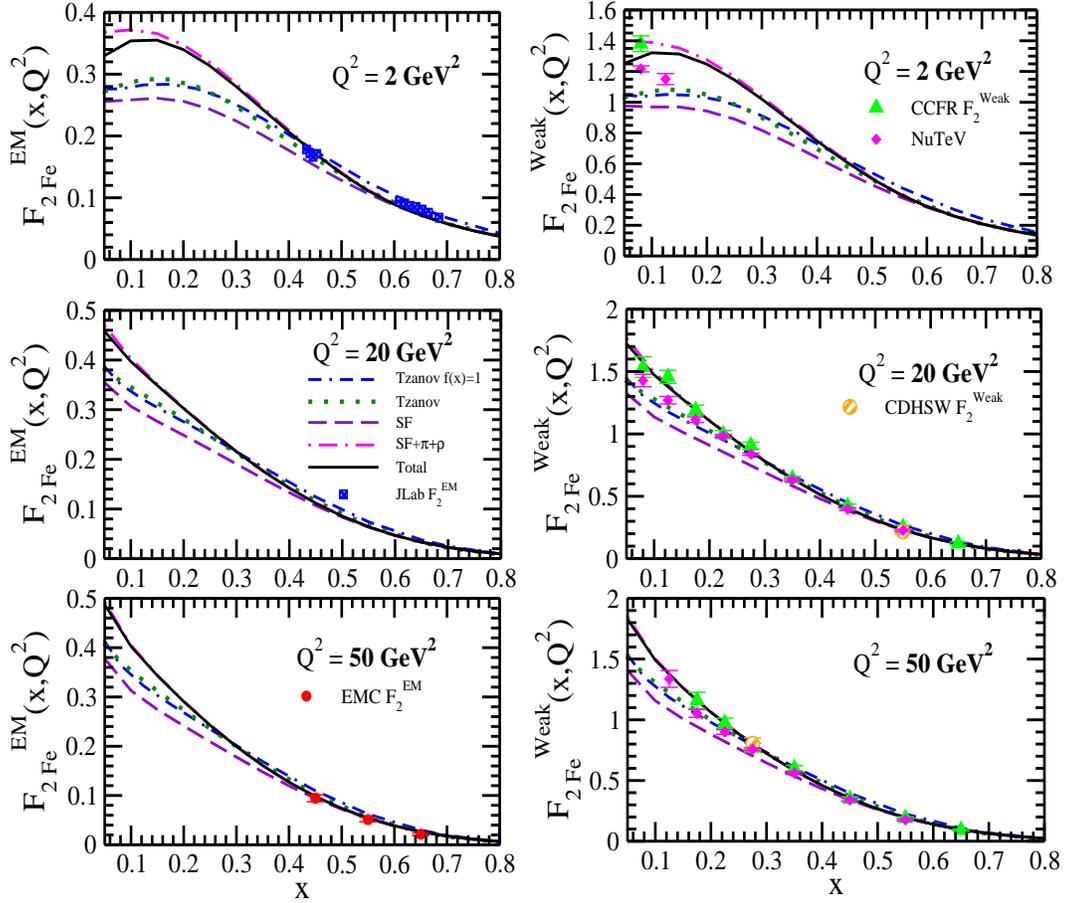}
 \caption{Results of EM(Left panel) and Weak(Right panel) nuclear structure functions in $^{56}Fe$(isoscalar) obtained 
 using spectral function(long dashed line), including mesonic 
contribution(dashed-dotted line), full model(solid line). The results are also presented with 
CCFR prescription i.e. $F_{2A}^{EM}(x,Q^2)=f(x) F_{2N}^{EM}(x,Q^2)$, 
 where $f(x)$~\cite{Seligman:1997fe}, is given by Eq.\ref{ggfx}(dotted line) and for the free nucleon case when $f(x)=1$(double dashed-dotted line).
These results are presented for different $Q^{2}$ and are compared with the available 
 experimental data of EMC(solid circle)~\cite{Aubert:1986yn}, JLab(square)~\cite{Mamyan:2012th},
 CDHSW(semi solid circle)~\cite{Berge:1989hr}, CCFR(triangle up)~\cite{Oltman:1992pq}, NuTeV(diamond)~\cite{Tzanov:2005kr}.}
 \label{fig0}
\end{figure}
\begin{figure}
\includegraphics[height=0.22\textheight,width=0.42\textwidth]{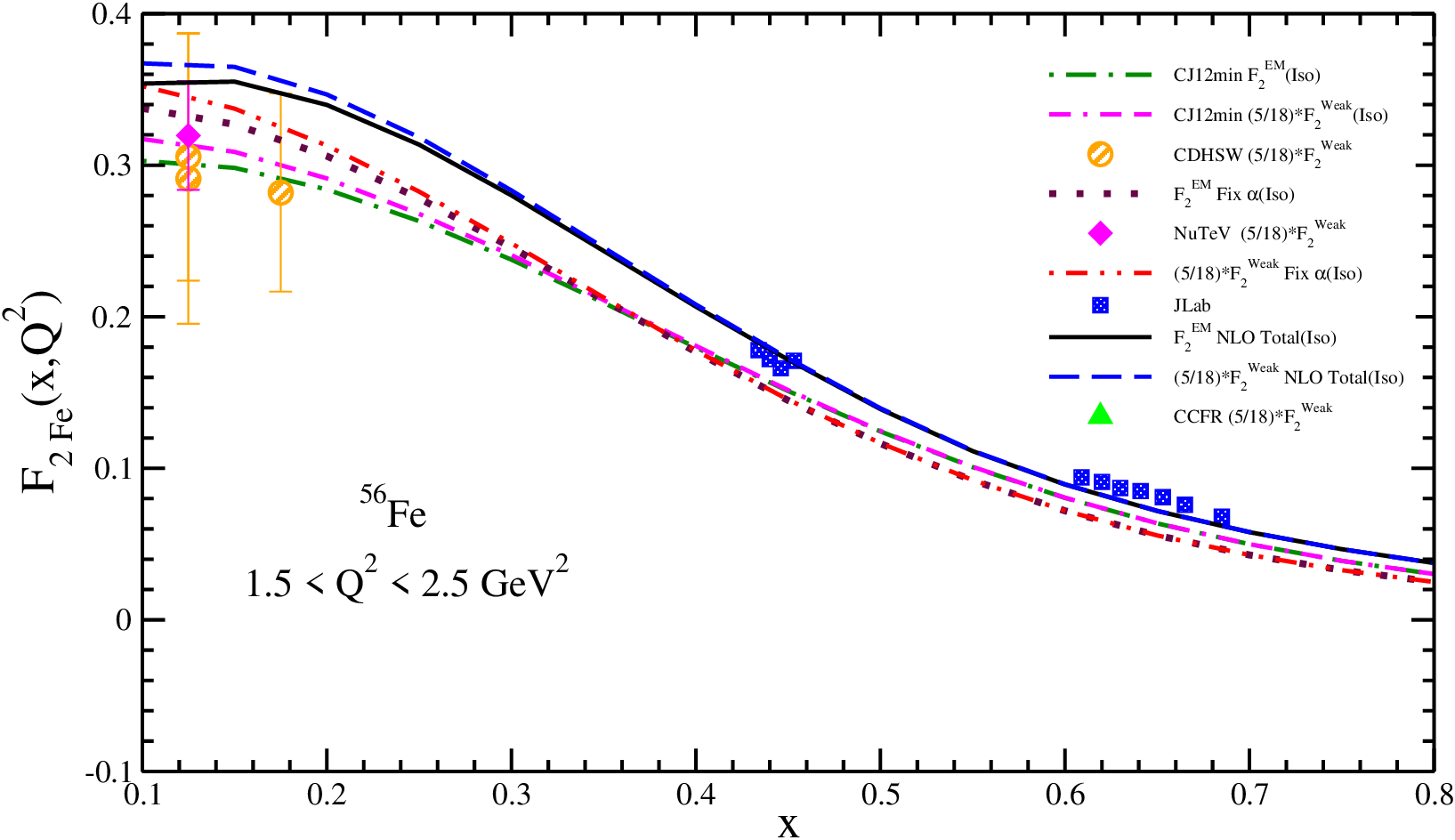}
\includegraphics[height=0.22\textheight,width=0.42\textwidth]{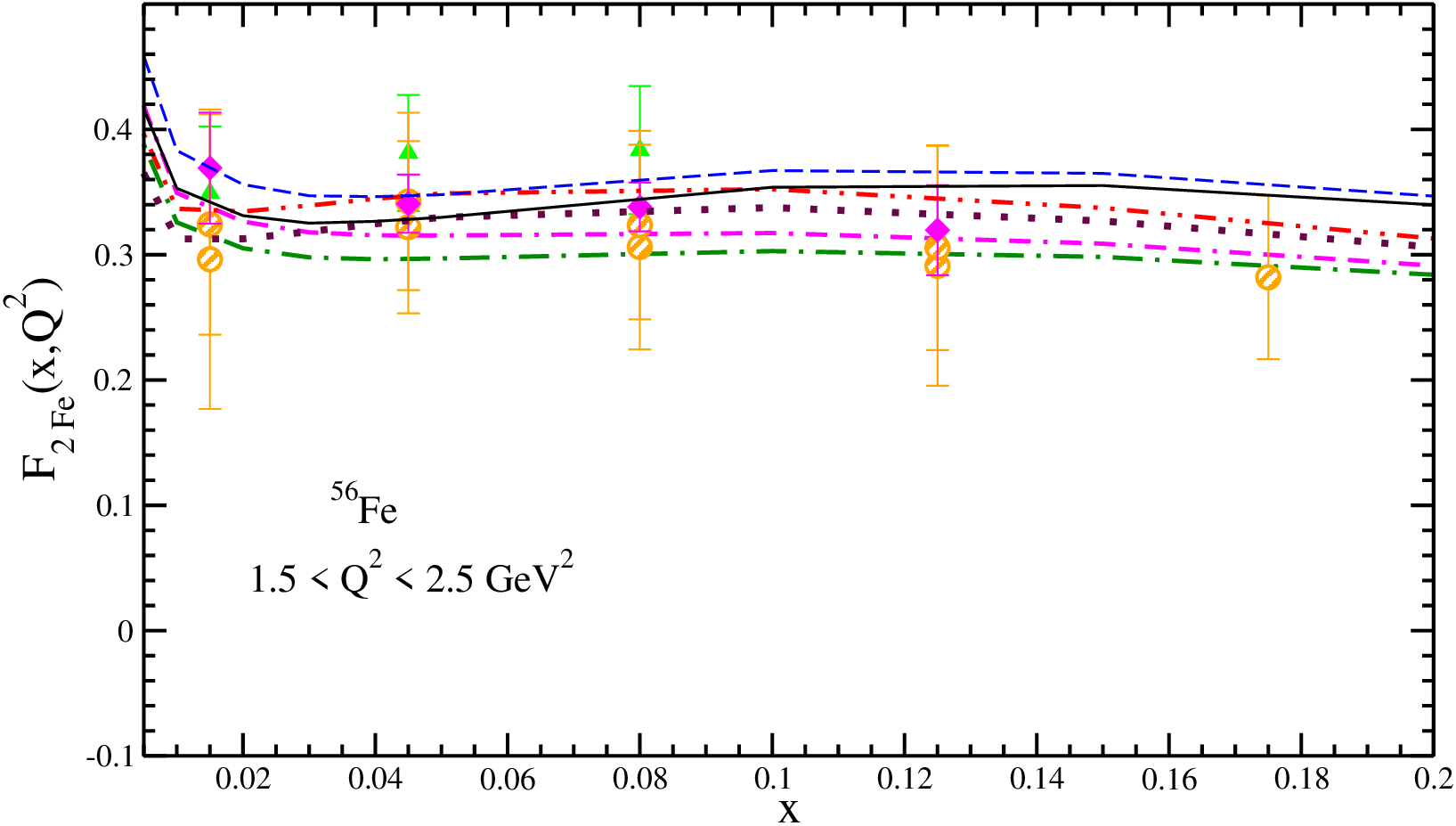}
\includegraphics[height=0.22\textheight,width=0.42\textwidth]{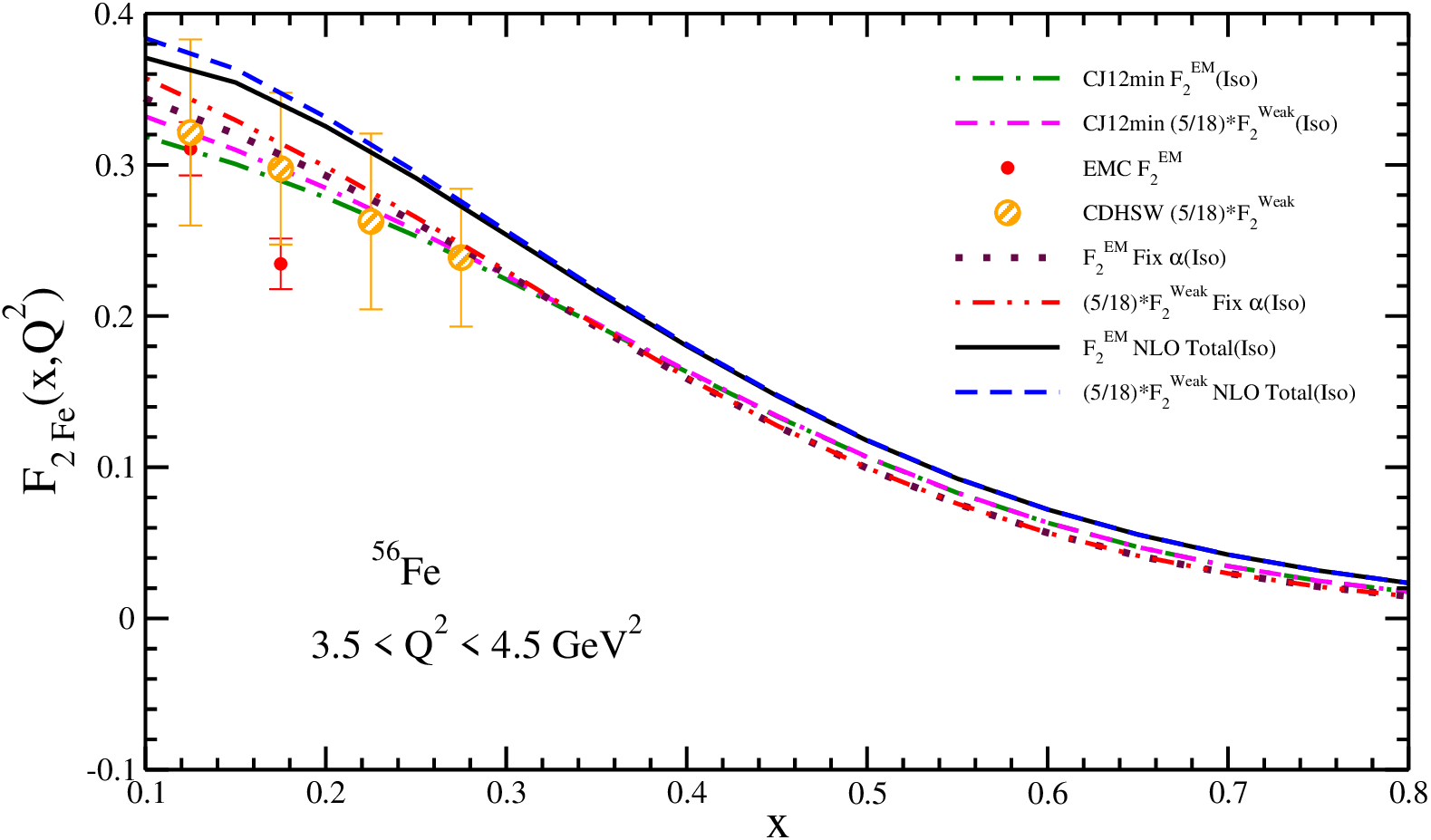}
\includegraphics[height=0.22\textheight,width=0.42\textwidth]{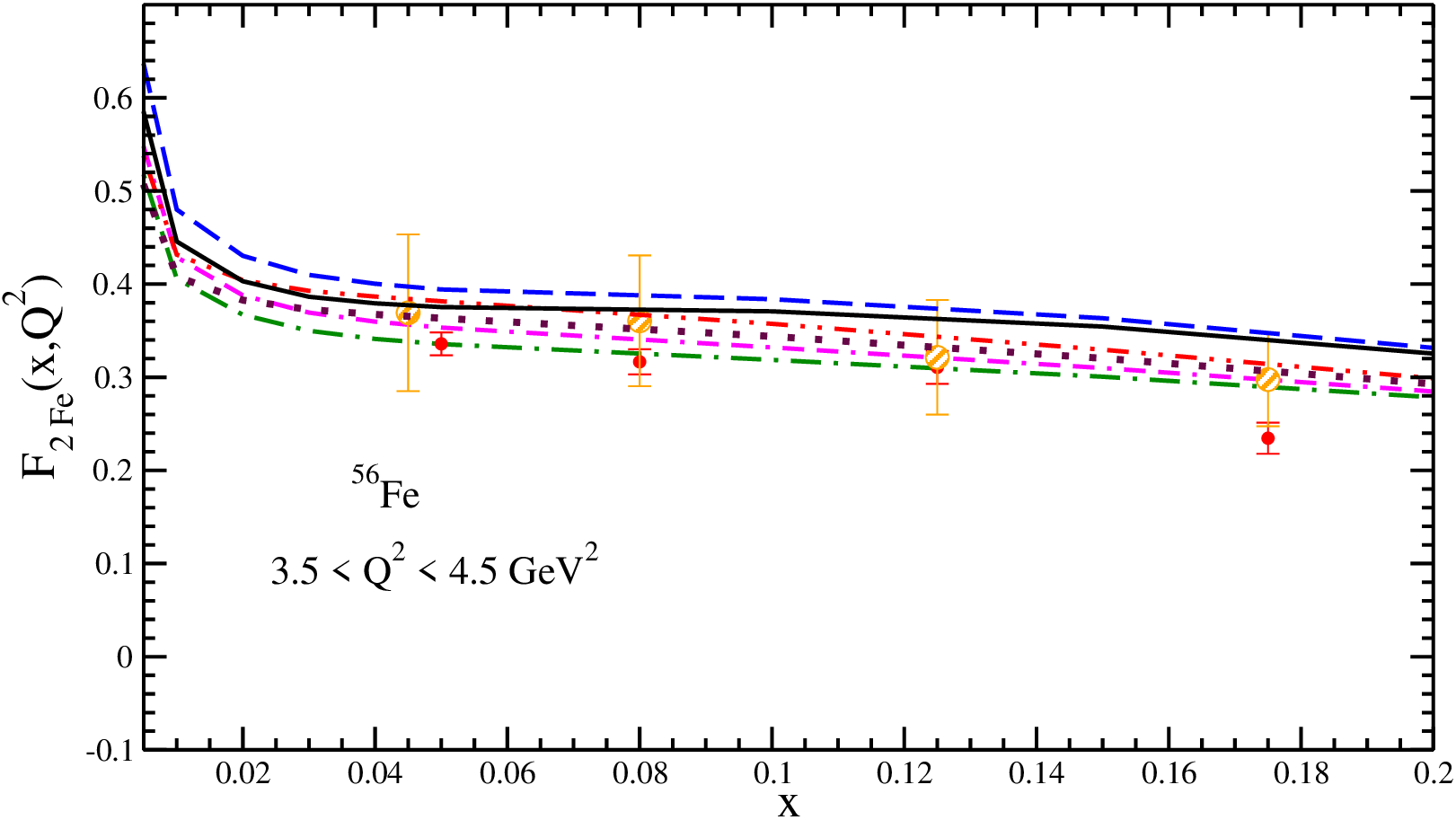}
\includegraphics[height=0.22\textheight,width=0.42\textwidth]{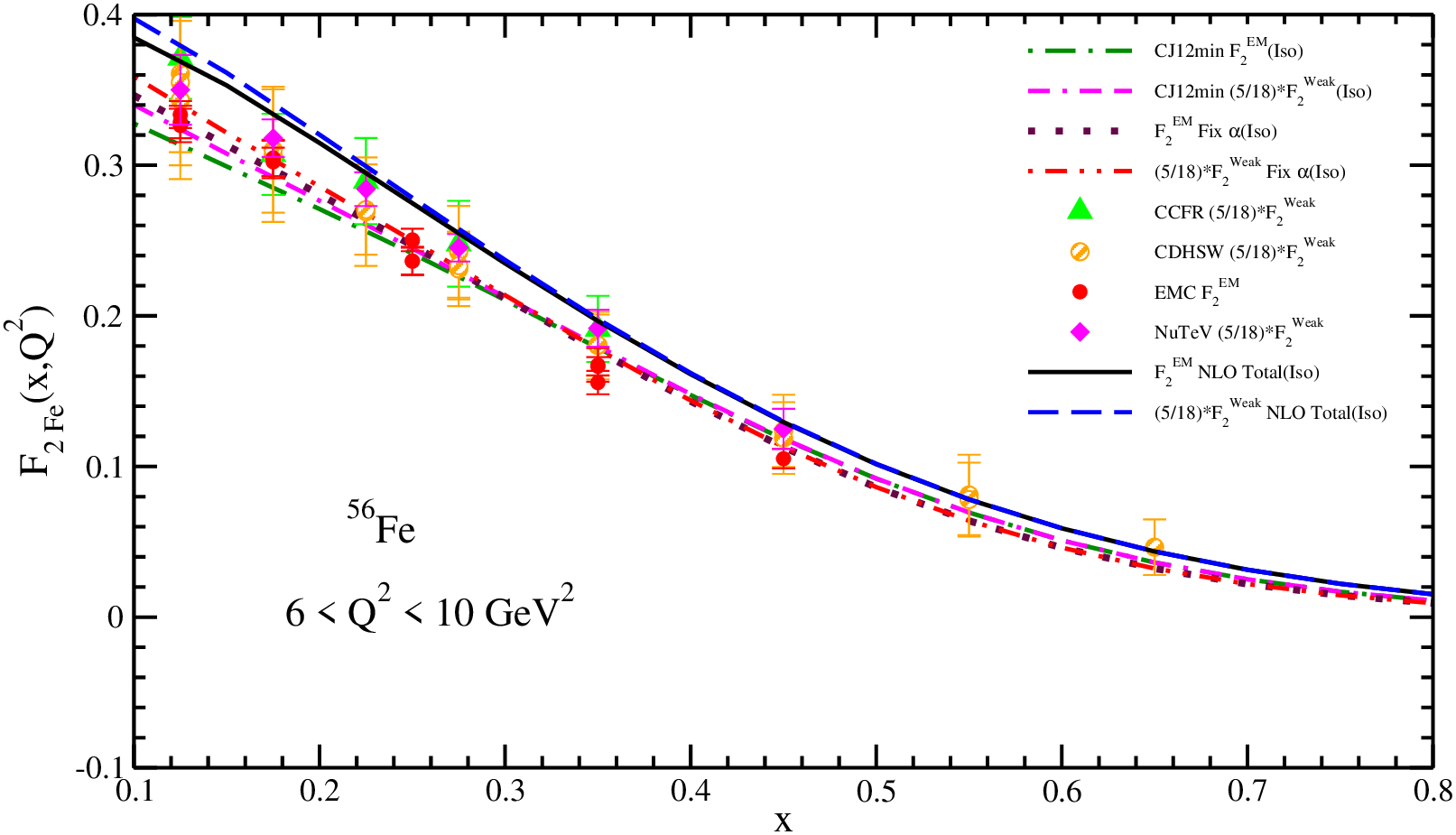}
\includegraphics[height=0.22\textheight,width=0.42\textwidth]{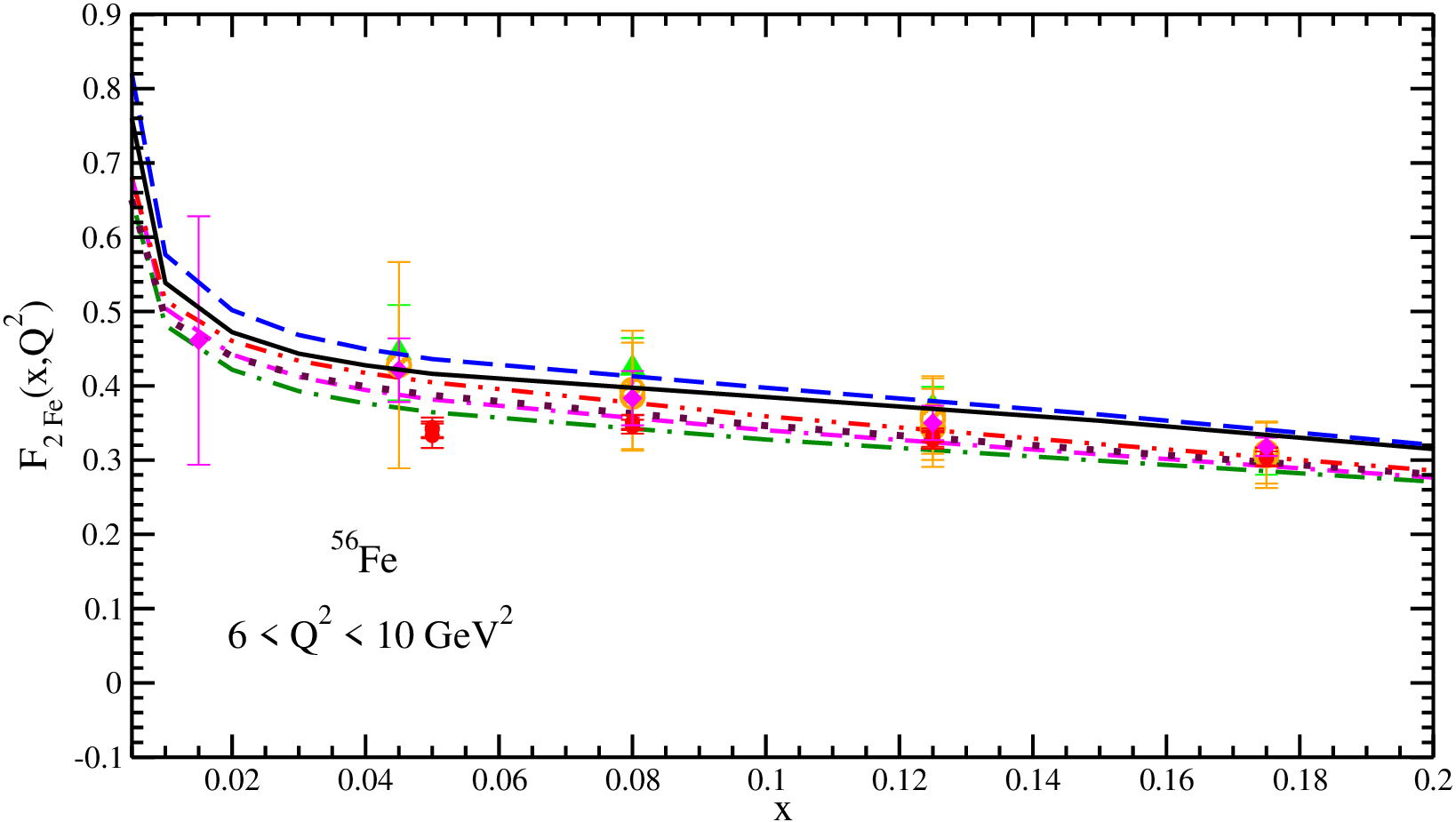}
\caption{Results of EM and Weak nuclear structure functions in $^{56}Fe$ are obtained at NLO using the full model(solid line and dashed line, respectively)
and also from CJ12min PDFs~\cite{Owens:2012bv} for fixed value of strong coupling constant(dotted line and dashed double dotted line, respectively) as well 
as for $Q^2$ evolution(dashed dotted line and double dashed dotted line, respectively).  The results of $F_2^{Weak}(x,Q^2)$ are scaled by a factor of $\frac{5}{18}$. The results are 
also compared with the available experimental data  EMC(solid circle)~\cite{Aubert:1986yn}, JLab(square)~\cite{Mamyan:2012th},
 CDHSW(semi solid circle)~\cite{Berge:1989hr}, CCFR(triangle up)~\cite{Oltman:1992pq}, NuTeV(diamond)~\cite{Tzanov:2005kr}. }
\label{fig1}
\end{figure}
\begin{figure}
\includegraphics[height=0.25\textheight,width=0.42\textwidth]{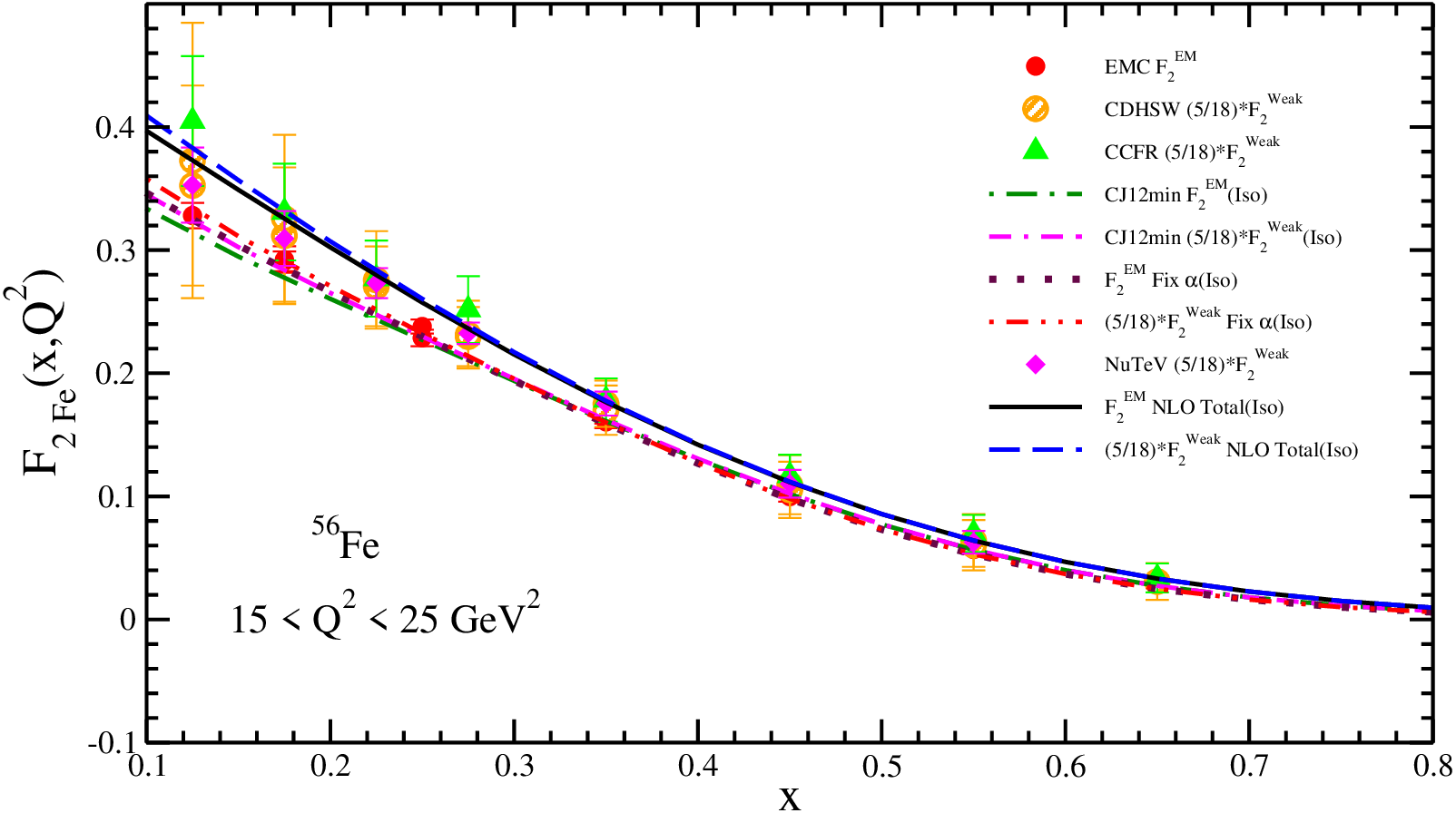}
\includegraphics[height=0.25\textheight,width=0.42\textwidth]{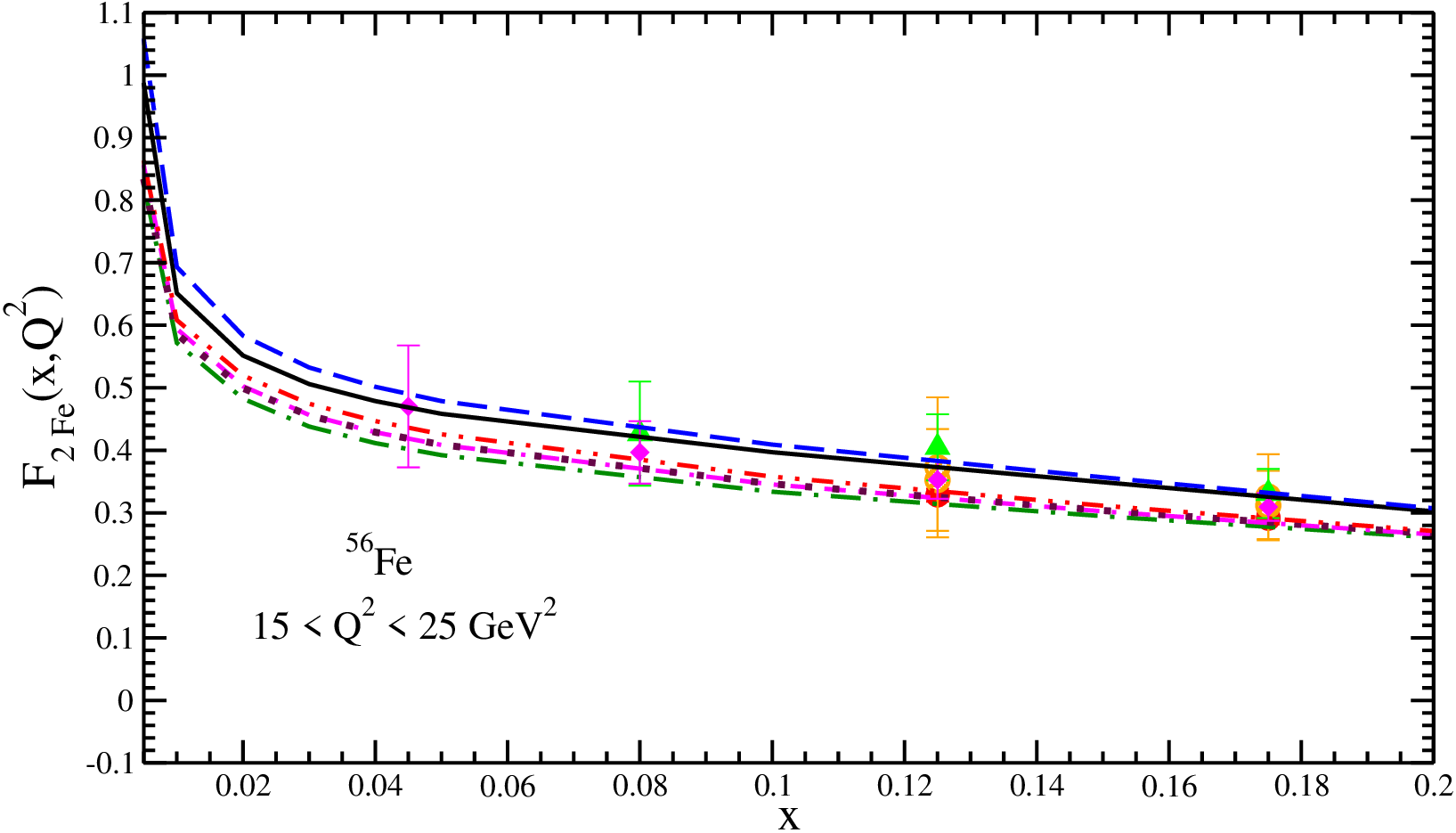}
\includegraphics[height=0.25\textheight,width=0.42\textwidth]{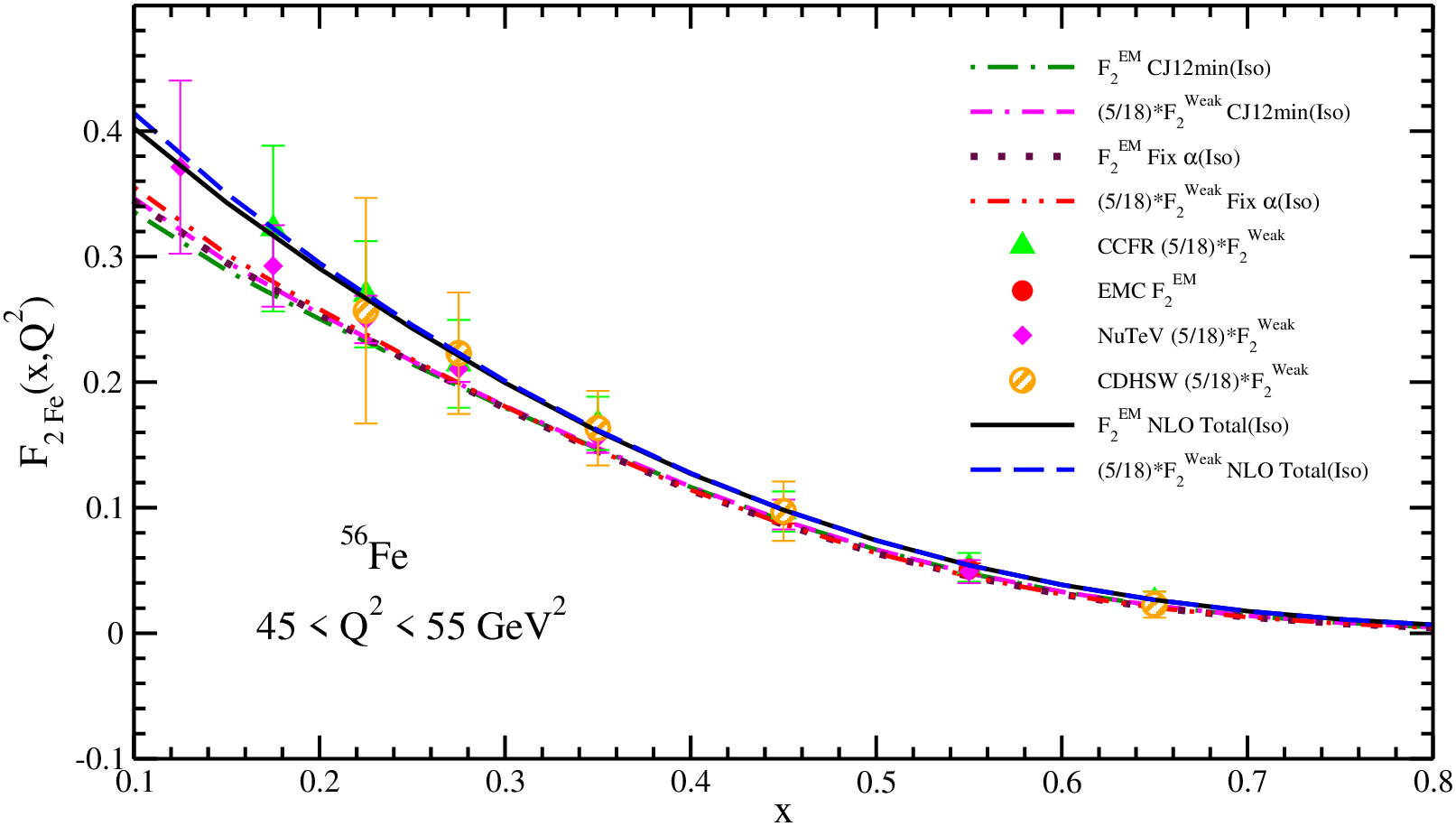}
\includegraphics[height=0.25\textheight,width=0.42\textwidth]{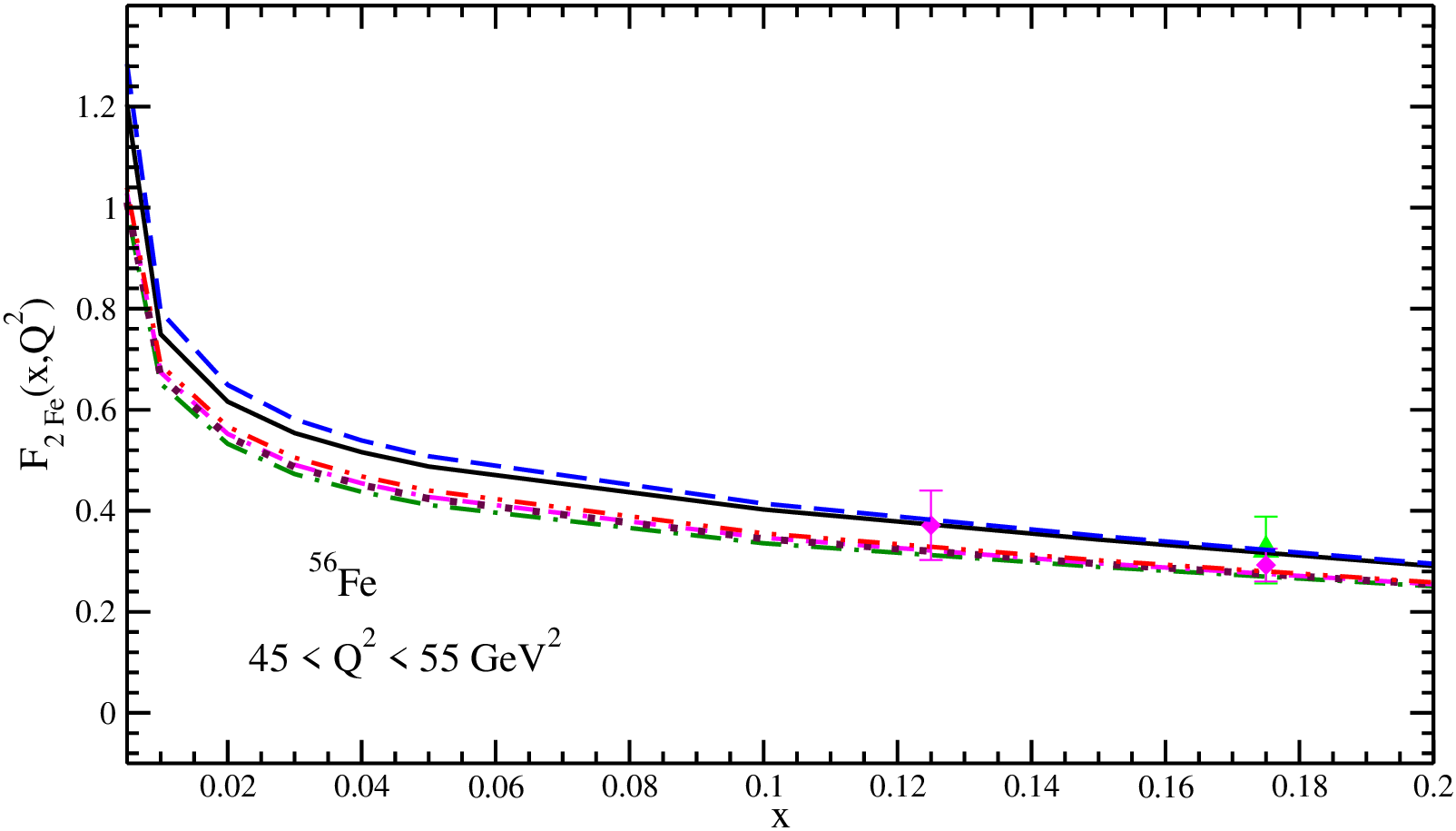}
\caption{Results of EM and Weak nuclear structure functions in $^{56}Fe$ obtained at NLO using the full model(solid line and dashed line, respectively)
and also from CJ12min PDFs~\cite{Owens:2012bv} for fixed value of strong coupling constant(dotted line and dashed double dotted line, respectively) as well 
as for $Q^2$ evolution(dashed dotted line and double dashed dotted line, respectively).  The results of $F_2^{Weak}(x,Q^2)$ are scaled by a factor of $\frac{5}{18}$. The results are 
also compared with the available experimental data  EMC(solid circle)~\cite{Aubert:1986yn}, 
 CDHSW(semi solid circle)~\cite{Berge:1989hr}, CCFR(triangle up)~\cite{Oltman:1992pq}, NuTeV(diamond)~\cite{Tzanov:2005kr}. }
\label{fig2}
\end{figure}
 \subsubsection{Mesonic effect}
  We have also taken into account virtual meson contribution in nuclear medium. This arises due to the presence of strong 
 attractive nature of nucleon-nucleon interactions, which in turn leads 
 to an increase in the interaction probability of virtual boson with the meson clouds. 
 To take into account mesonic effect, we have used microscopic approach
 by making use of the imaginary part of the meson propagators instead of spectral function, and obtain
 $F_{2 A,\pi}(x,Q^2)$~\cite{Marco:1995vb} as
\begin{eqnarray} \label{pion_f21}
F_{_{2 A,\pi}}^{ EM, Weak}(x,Q^2)  &=&  -6 \int \, d^3 r \, \int \frac{d^4 p}{(2 \pi)^4} \, 
        \theta (p^0) ~\delta I m D (p) \;2m_\pi~\left(\frac{m_\pi}{p^0~-~p^z~\gamma}\right)\times \nonumber \\
&&\left[\frac{Q^2}{(q^z)^2}\left( \frac{|{\bf p}|^2~-~(p^{z})^2}{2m_\pi^2}\right)  
+  \frac{(p^0~-~p^z~\gamma)^2}{m_\pi^2} \left(\frac{p^z~Q^2}{(p^0~-~p^z~\gamma) q^0 q^z}~+~1\right)^2\right] ~F_{_{2\pi}}^{ EM, Weak}(x_\pi)~~
\end{eqnarray}
where $x_\pi=-\frac{Q^2}{2p \cdot q}$, $m_\pi$ is the pion mass and $D(p)$ is the pion propagator in the nuclear medium given by 
 \begin{equation}\label{dpi}
D (p) = [ p_0^2 - {\bf {p}}\,^{2} - m^2_{\pi} - \Pi_{\pi} (p_0, {\bf p}) ]^{- 1}\,,
\end{equation}
where
\begin{equation}\label{pionSelfenergy}
\Pi_\pi=\frac{f^2/m_\pi^2 F^2(p){\bf {p}}\,^{2}\Pi^*}{1-f^2/m_\pi^2 V'_L\Pi^*}\,.
\end{equation}
Here, $F(p)=(\Lambda_\pi^2-m_\pi^2)/(\Lambda_\pi^2 - p^2)$ is the $\pi NN$ form factor, $p^2=p_0^2~-~{\bf p}^2$, $\Lambda_\pi$=1~$GeV$, $f=1.01$, $V'_L$ is
the longitudinal part of the spin-isospin interaction and $\Pi^*$ is the irreducible pion self energy that contains the contribution of
particle - hole and delta - hole excitations.

Similarly the contribution of the $\rho$-meson cloud to the structure function is taken into account in analogy with the above model and the
rho structure function is written as~\cite{Marco:1995vb}.
\begin{eqnarray} \label{F2rho1}
F_{_{2 A,\rho}}^{ EM, Weak}(x,Q^2)  &=& -12 \int \, d^3 r \, \int \frac{d^4 p}{(2 \pi)^4} \, 
        \theta (p^0) ~\delta I m D_\rho (p) \;2m_\rho~\left(\frac{m_\rho}{p^0~-~p^z~\gamma}\right) \times \nonumber \\
&&\left[\frac{Q^2}{(q^z)^2}\left( \frac{|{\bf p}|^2~-~(p^{z})^2}{2m_\rho^2}\right)  
+  \frac{(p^0~-~p^z~\gamma)^2}{m_\rho^2} \left(\frac{p^z~Q^2}{(p^0~-~p^z~\gamma) q^0 q^z}~+~1\right)^2\right]~F_{_{2\rho}}^{ EM, Weak}(x_\rho)~~~~
\end{eqnarray}
where $x_\rho=-\frac{Q^2}{2p \cdot q}$, $m_\rho$ is the mass of $\rho$ meson and $D_{\rho} (p)$ is now the $\rho$-meson propagator in the medium given by:
\begin{equation}\label{dro}
D_{\rho} (p) = [ p_0^2 - {\bf {p}}\,^{2} - m^2_{\rho} - \Pi^*_{\rho} (p_0, {\bf p}) ]^{- 1}\,,
\end{equation}
where
\begin{equation}\label{rhoSelfenergy}
\Pi^*_\rho=\frac{f^2/m_\pi^2 C_\rho F_\rho^2(p){\bf{p}}\,^{2}\Pi^*}{1-f^2/m_\pi^2 V'_T\Pi^*}\,.
\end{equation}
Here, $V'_T$ is the transverse part of the spin-isospin interaction, $C_\rho=3.94$, $F_\rho(p)=(\Lambda_\rho^2-m_\pi^2)/(\Lambda_\rho^2 - p^2)$ is the $\rho NN$ form factor, 
$\Lambda_\rho$=1~$GeV$, $f=1.01$, and $\Pi^*$ is the irreducible rho self energy that contains the contribution of particle - hole and delta - hole excitations.
Quark and antiquark PDFs for pions have been taken from the parameterization given by Gluck et al.\cite{Gluck:1991ey} and for the 
rho mesons we have taken the same PDFs as for the pions.

It should be emphasized that the choice of $\Lambda_\pi$ and $\Lambda_\rho=$ 1 GeV respectively used in the expressions of $\pi NN$ and $\rho NN$ form factors
  given in Eqs.~\ref{pionSelfenergy} and \ref{rhoSelfenergy}, have been fixed in our earlier works~\cite{sajjadnpa, Haider:2015vea} while describing nuclear
medium effects in electromagnetic nuclear structure function $F_2^{EM}(x,Q^2)$ to explain the latest data from JLab and 
other experiments performed using charged lepton beam scattering from several nuclear targets in the DIS region.
  However, in literature, the value of $\Lambda_\pi$ and $\Lambda_\rho$ in the range of 1-1.3~GeV are used. We have also studied the uncertainty in 
  $F_2^{EM}(x,Q^2)$ and $F_2^{Weak}(x,Q^2)$ arising due to the use of different cut off parameters ($\Lambda_\pi$ and $\Lambda_\rho$) by varying them in the region of 0.9-1.1~GeV i.e. 
  10$\%$ variation from the central value of 1GeV. 
\begin{figure}
 \includegraphics[height=0.45\textheight,width=0.8\textwidth]{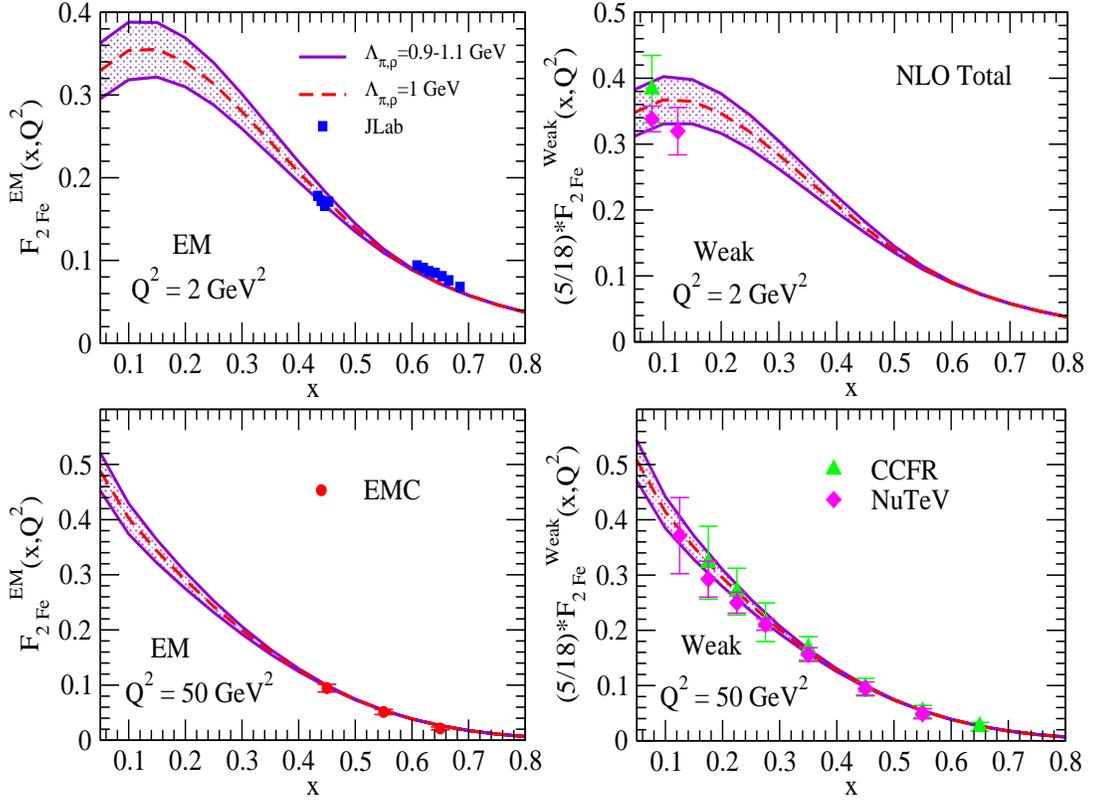}
 \caption{Results for EM(Left panel) and weak(Right panel) nuclear structure functions $F_{2A}(x,Q^2)$  are shown for $A = ~^{56}Fe$(isoscalar) at different values of $Q^2=2,50~GeV^2$.
Effect of the parameters $\Lambda_\pi$ and $\Lambda_\rho$ used in the expressions of $\pi NN$ and $\rho NN$ form factors given in
Eqs.\ref{pionSelfenergy}, \ref{rhoSelfenergy} respectively, that are varied by 10$\%$ from the central value is shown.
% in the meson cloud contribution. 
The results are obtained with full model at NLO and are compared with the 
 experimental data of EMC(solid circle)~\cite{Aubert:1986yn}, JLab(square)~\cite{Mamyan:2012th}, CCFR(triangle up)~\cite{Oltman:1992pq}, NuTeV(diamond)~\cite{Tzanov:2005kr}.}
 \label{meson_fig}
\end{figure}

For the shadowing effect, we are following the model of Kulagin and Petti~\cite{Kulagin:2004ie} who have 
used Glauber-Gribov multiple scattering theory.
By using the above expression(Eq.~\ref{F2rho1}) for weak and electromagnetic nuclear structure functions we have performed the
 calculations in carbon, aluminum, calcium, iron, copper, tin and lead nuclei and
the results are presented for different kinematic regions of $x$ and $Q^2$.

\begin{figure}
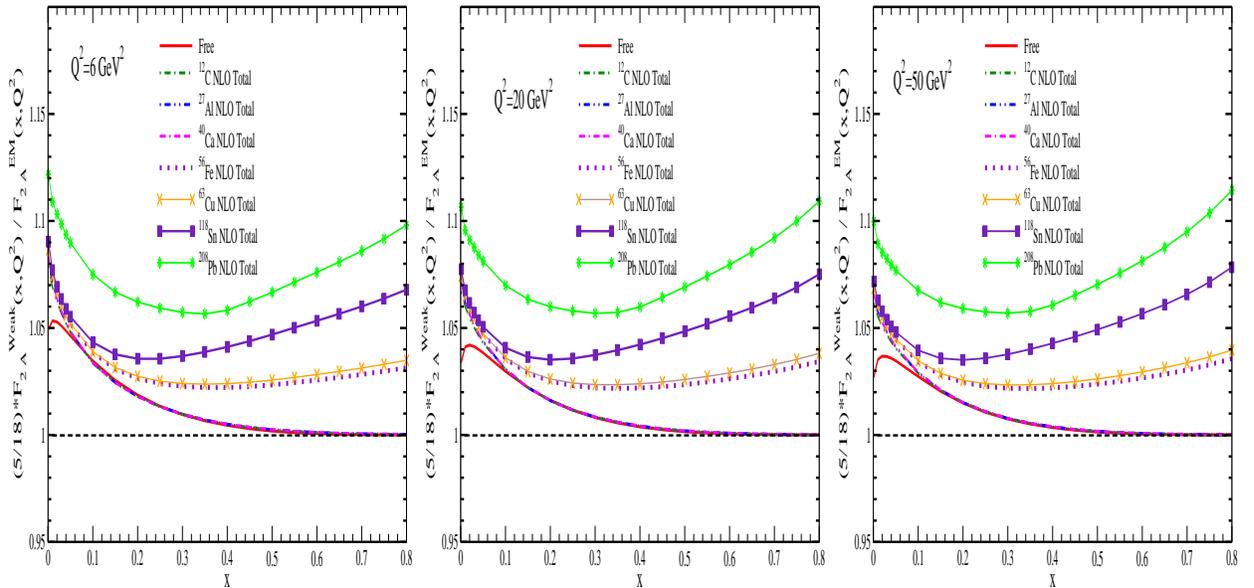

\includegraphics[height=0.33\textheight,width=0.3\textwidth]{calcafecusnpb_ew_6q2.eps}
\includegraphics[height=0.33\textheight,width=0.3\textwidth]{calcafecusnpb_ew_20q2.eps}
\includegraphics[height=0.33\textheight,width=0.3\textwidth]{calcafecusnpb_ew_50q2.eps}
\caption{Results for the ratio ${\it R^\prime}=\frac{\frac{5}{18} F_{2 A}^{Weak}(x,Q^2)}{F_{2 A}^{EM}(x,Q^2)}$ are obtained
by using the full model at NLO 
 in $A=$ $^{12}C$, $^{27}Al$, $^{40}Ca$, $^{56}Fe$, $^{63}Cu$, $^{118}Sn$ and $^{208}Pb$ at $Q^2=6,~20,~50~GeV^2$.}
\label{fig4}
\end{figure}

\begin{figure}
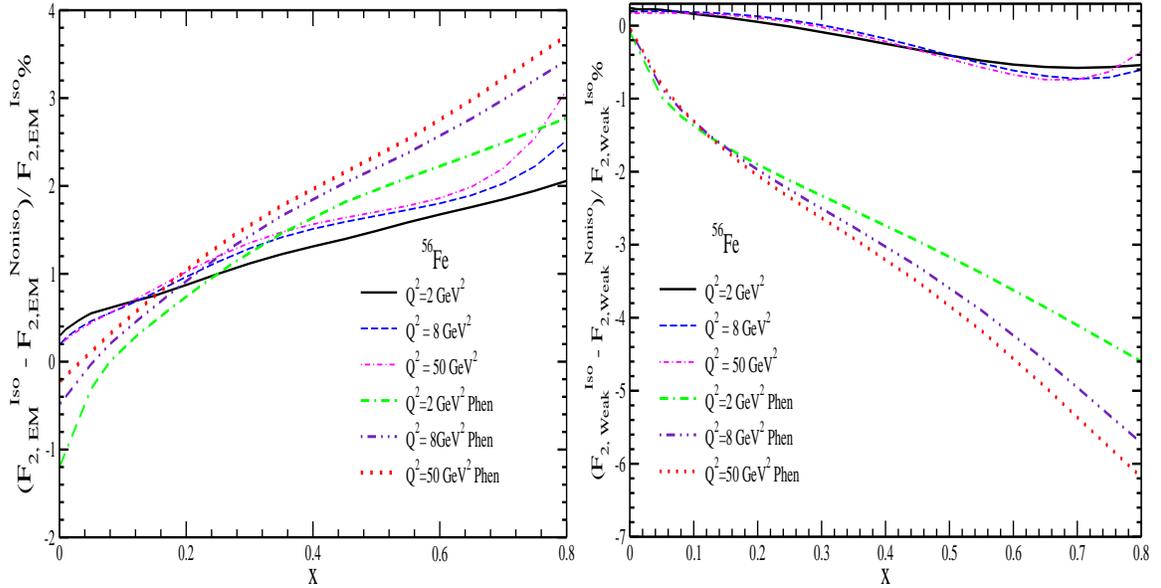

\includegraphics[height=0.33\textheight,width=0.42\textwidth]{iso_noniso_em.eps}
\includegraphics[height=0.33\textheight,width=0.42\textwidth]{iso_noniso_weak.eps}
\caption{$\%$ deviation from Isoscalarity (i.e. $\frac{F_2^{Iso}~-~F_2^{NonIso}}{F_2^{Iso}}$) in $^{56}Fe$ for EM(Left) 
and Weak(Right) nuclear structure functions at different $Q^2$. Solid(double dashed dotted) 
line is the result at $Q^2=2~GeV^2$, long dashed(dashed double dotted) line is the 
result at $Q^2=8~GeV^2$ and
short dashed dotted(dotted) line is the result at $Q^2=50~GeV^2$ 
obtained by using the full model(phenomenological prescription using 
Eqs. \ref{gg1}, \ref{gg2}) at NLO.}
\label{fig5}
\end{figure}

\begin{figure}
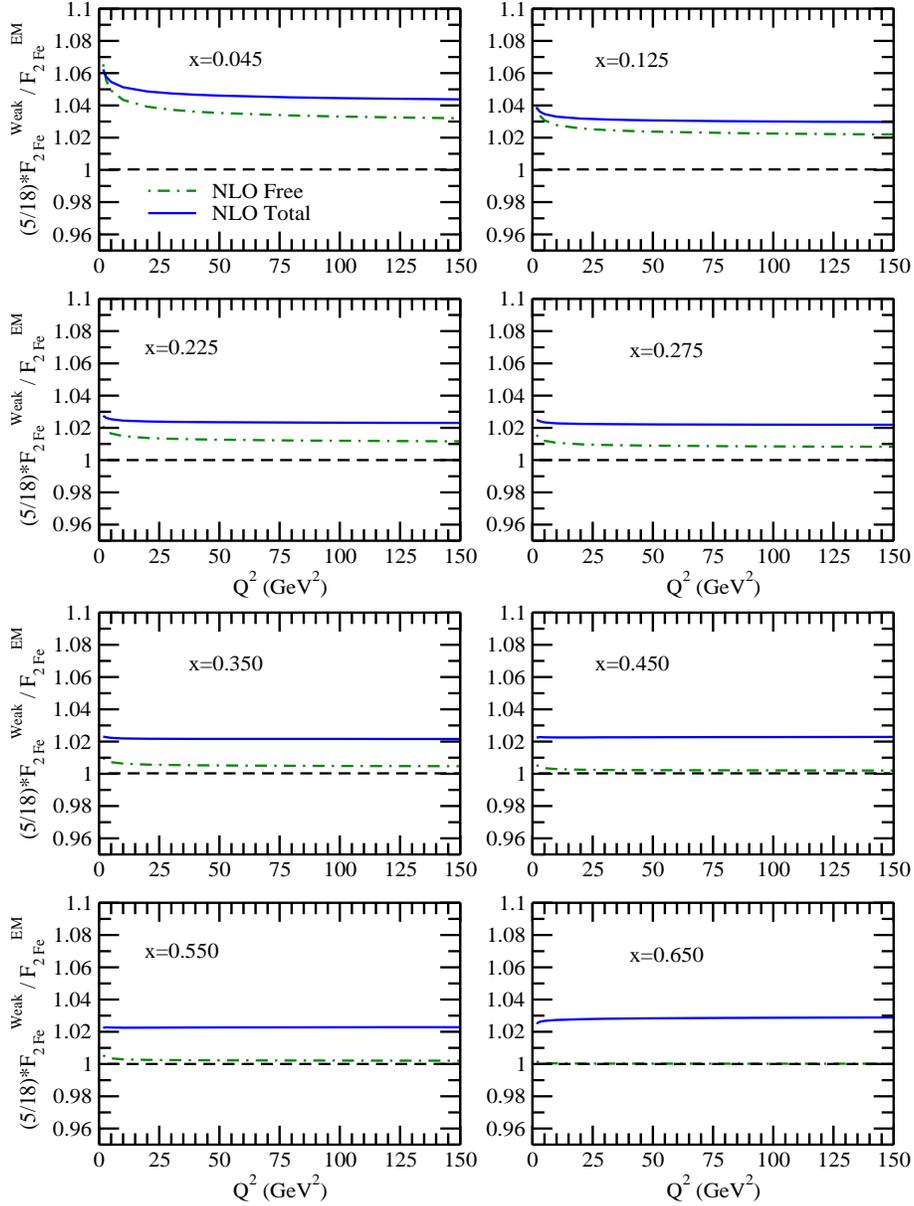

\includegraphics[height=8 cm , width= 12 cm]{nlo_ratiio1.eps}
\includegraphics[height=8 cm , width= 12 cm]{nlo_ratiio2.eps}
\caption{Results for the ratio ${\it R^\prime}=\frac{\frac{5}{18} F_{2~Fe}^{Weak}(x,Q^2)}{F_{2~Fe}^{EM}(x,Q^2)}$ vs $Q^2$ 
are shown at NLO in $^{56}Fe$ for different values of Bjorken scaling variable $x$. Solid(dashed-dotted)
line is result obtained by using the full model(free nucleon case) at NLO.}
\label{fig6}
\end{figure}

\section{Results and Discussion}\label{sec:RE}
We have considered Fermi motion, binding energy, nucleon correlations using spectral function to evaluate the structure functions, given in 
Eqs.~\ref{had_ten151} and \ref{had_aten151} for isoscalar and nonisoscalar nuclear targets, respectively. The mesonic contributions from pion and rho mesons are calculated 
using Eq.~\ref{pion_f21} and Eq.~\ref{F2rho1} respectively, while shadowing effects are included using the work of Kulagin and Petti~\cite{Kulagin:2004ie}. 
 PDFs parameterized by CTEQ group~\cite{Nadolsky:2008zw} have been used for numerical calculations 
 and the evaluations are made at the Next-to-Leading-Order(NLO). This is our full model.

To compare the present results for the structure functions $F_{2A}^{Weak}(x,Q^2)$ and $F_{2A}^{EM}(x,Q^2)$ with the results of CTEQ collaboration~\cite{cteqcolla}, 
 we have used their CJ12min PDFs set~\cite{Owens:2012bv}. The calculations are performed for 
fixed values of strong coupling constant 
($\alpha_s(M_Z)=0.118$ the approximate world average value) and the results are presented along with the present results obtained using the microscopic model.
 
 In Fig.~\ref{fig0}, we present the results for $F_{2A}^{EM}(x,Q^2)$ vs $x$ and $F_{2A}^{Weak}(x,Q^2)$ vs $x$ at 
 different values of $Q^2$ viz. 2, 20, 50 $GeV^2$. 
 In this figure, we show the curves for $F_{2A}^{EM,Weak}(x,Q^2)$ obtained using the spectral
 function only, also including the mesonic contribution, 
 and finally using the full model which also includes shadowing. 
 The use of nuclear spectral function which takes into account Fermi motion,
 binding energy and nucleon correlations, leads to a reduction of $\sim 8\%$ at $x=0.1$; $\sim 18\%$ at $x=0.4$; $\sim 3\%$ at $x=0.7$
 in $F_{2A}^{EM}(x,Q^2)$ as well as in $F_{2A}^{Weak}(x,Q^2)$ nuclear structure functions as compared to the free nucleon case. 
 The inclusion of mesonic contributions from pion and rho mesons leads to an enhancement in these structure functions at low and medium values of $x$. For example, 
  the enhancement is $\sim 30\%$ at $x$=0.1; $\sim 15\%$ at $x$=0.4; $\sim 0.3\%$ at $x$=0.7. The inclusion of shadowing effects further 
   reduces these structure functions and are effective in low region of $x~(~<~0.1)$. For example, 
  the reduction is $\sim 10\%$ at $x$=0.05 and $\sim 5\%$ at $x$=0.1.
  Thus, we find that the mesonic contributions are sizable in the low and medium values of $x$, 
  while the shadowing effects are non-negligible only in the low region of $x$.
    Therefore, when mesonic and shadowing effects are taken along with the nuclear medium effects obtained by using the spectral function 
  there is an overall change in the numerical results from the free nucleon values which lead to better explanation of experimental data.
      
  Here we have also shown the results obtained by using the prescription adopted by CCFR group~\cite{Seligman:1997fe}, 
 where nuclear correction factor $f(x)$ is obtained by analyzing charged lepton scattering data on isoscalar 
 nuclear targets. In these analyses nonisoscalar
  nuclear targets have also been considered after making isoscalarity correction.
  The nuclear structure functions are then expressed as a product of free nucleon structure functions and nuclear correction factor i.e.
 \begin{eqnarray}\label{gg}
 F_{2A}^{EM}(x,Q^2)&=&f(x) F_{2N}^{EM}(x,Q^2)\nonumber \\ 
 F_{2A}^{Weak}(x,Q^2)&=&f(x)  F_{2N}^{Weak}(x,Q^2)
 \end{eqnarray}
 where $f(x)$ is defined in Eq.~\ref{ggfx}. We have also shown in this figure the 
 results for isoscalar nucleon case.
 
 In Figs.~\ref{fig1} and ~\ref{fig2}, we present the results for nuclear structure functions
 $F_{2A}^{EM,Weak}(x,Q^2)$ for electromagnetic and weak interactions in 
 iron nucleus in the $Q^2$ region $1.5 ~<~ Q^2 ~<~10~GeV^2$ and $15 ~<~ Q^2 ~<~55~GeV^2$ respectively, 
 using the full model at NLO.
 These results are compared with the available experimental data from the experiments performed with charged 
 lepton beams at JLab~\cite{Mamyan:2012th}, 
 EMC~\cite{Aubert:1986yn}, SLAC~\cite{Gomez:1993ri, Hen:2013oha, Whitlow:1990dr, Dasu:1993vk} and neutrinos/antineutrinos data from
  CDHSW~\cite{Berge:1989hr}, CCFR~\cite{Oltman:1992pq}, NuTeV~\cite{Tzanov:2005kr} measurements. The theoretical results of structure functions shown 
  in various panels of Figs.~\ref{fig1} and ~\ref{fig2} correspond to the central values of the range of $Q^2$ in these figures. 
  We see that the present results are consistent with  CCFR~\cite{Oltman:1992pq},
 JLab~\cite{Mamyan:2012th}, NuTeV~\cite{Tzanov:2005kr}
 data at medium and high values of $Q^2$ but not in good agreement at 
 low $Q^2$ with the experiments like CDHSW~\cite{Berge:1989hr} and EMC~\cite{Aubert:1986yn}.
 From these figures, one may also observe that at low $x$, EM structure 
 function is slightly lower than the weak structure function which is about $\sim 4\%$ in iron 
 at $x=0.1$, and the difference decreases with 
  the increase in $x$ and almost becomes negligible for $x~>~0.3$. We find that(not shown here) 
  with the increase in mass number this difference increases. For example, in lead it becomes $\sim 7\%$ 
  while for low mass nuclei like carbon this difference 
  becomes $\sim 1-2\%$ at $x=0.1$.
  Here we also compare our results with the results obtained by CTEQ collaboration~\cite{Nadolsky:2008zw} using
  CJ12min PDFs~\cite{Owens:2012bv}. These effects could be a hint of the importance of heavier quark flavous in large mass nuclei.

In Fig.\ref{meson_fig}, we show the uncertainty in $F_{2A}^{EM}(x,Q^2)$ and $F_{2A}^{Weak}(x,Q^2)$ by changing the parameters $\Lambda_\pi$ and $\Lambda_\rho$ 
 by 10$\%$ around the central value of 1GeV, and the results are shown by putting a band.  
 It is seen that at low value of $Q^2$($2GeV^2$), there is significant dependence of $F_{2A}(x,Q^2)$ on the choice of $\Lambda_\pi$ 
 and $\Lambda_\rho$, for example the deviation is 
 $\sim 8-10\%$ at low values of $x$(=0.1), $\sim 6\%$ at $x$=0.4 and almost negligible for $x>0.6$. 
 While for higher values of $Q^2$ this deviation is around $6-7\%$ at $x$=0.1, $2\%$ at $x$=0.4 and becomes negligible for large $x$.
  This variation has been found to be of similar in strength both for the electromagnetic 
 and weak structure functions and thus 
 is hardly going to change the ratio $\frac{F_2^{EM}(x,Q^2)}{\frac{5}{18}F_2^{Weak}(x,Q^2)}$ . 
%  To explicitly show the effect of the variation in $\Lambda_\pi$ 
%  and $\Lambda_\rho$ parameters on the ratio of nuclear structure functions, we have shown in Fig.\ref{expli}, 
%  $\frac{\frac{18}{5}F_{2 A}^{EM}(x,Q^2)}{F_{2 A}^{Weak}(x,Q^2)}$ at $Q^2=2 ~GeV^2$ and $50~GeV^2$ for the free nucleon case with and without sea contribution where there 
%  would be no nuclear effect for the free case, and performing calculations using the present model with full prescription with $\Lambda_{\pi,\rho}$=0.9~GeV,~1~GeV, and 1.1~GeV 
%  in Eqs. \ref{pion_f21}, \ref{F2rho1}, respectively. We find that it is mainly the sea quark contribution which makes the difference in the ratio.
%  The results are almost independent($<~1\%$) of the variation in the values of $\Lambda_\pi$ and $\Lambda_\rho$. 
  
 In Fig.~\ref{fig4}, we show the ratio $R^\prime=\frac{\frac{5}{18} F_{2A}^{Weak}(x,Q^2)}{F_{2A}^{EM}(x,Q^2)}$ in 
 various nuclei like 
 $^{12}C$, $^{27}Al$, $^{40}Ca$, $^{56}Fe$, $^{63}Cu$, $^{118}Sn$ and $^{208}Pb$ at $Q^2=6,~20,~50~GeV^2$.
 These results are obtained assuming the targets to be nonisoscalar wherever applicable. One may observe that for heavier nuclear targets like
 $^{118}Sn$ and $^{208}Pb$, the nonisoscalarity effect is larger. 
  This shows that the difference in charm and strange quark distributions could be
  significant in heavy nuclei. 
  Furthermore, we find an $x$ dependence as well as a $Q^2$ dependence in this ratio.
 
 In order to explicitly quantify the effect of nonisoscalarity, in Fig.~\ref{fig5}, we 
 plot the ratio $r(x,Q^2)=\frac{F_{2}^{Iso}(x,Q^2)~-~F_{2}^{NonIso}(x,Q^2)}{F_{2}^{Iso}(x,Q^2)}$ vs $x$, where $F_{2}^{Iso}(x,Q^2)$ is obtained 
 by using Eq.~\ref{had_ten151} and $F_{2}^{NonIso}(x,Q^2)$ is obtained by using Eq.~\ref{had_aten151} with
 full model at NLO, 
 by considering isoscalar as well as nonisoscalar nuclear targets, 
 at different values of $Q^2$ viz. 2, 8, 50 $GeV^2$,
 for electromagnetic and weak structure functions. The results are compared with the phenomenological prescription of $R_{ISO}^{A,EM}$ and $R_{ISO}^{A,Weak}$
 given in Eqs.\ref{gg1} and \ref{gg2}  for 
 electromagnetic and weak processes, respectively.
 In the present model for the electromagnetic case, the deviation from isoscalarity is less than 1$\%$ at smaller values of $x$, which increases to
 around $2-3\%$ at higher values of $x$. The present results are also compared with the results obtained by using 
 phenomenological prescription given in Eqs.\ref{gg1} and \ref{gg2}. We find that the difference between 
 the phenomenological approach and the present approach is less than a percent for all values of $x$.
 
 While in the weak interaction case, we find a larger difference in the results obtained by using phenomenological approach 
 and the present approach. Using the present model we find almost no difference in the results, whereas using Eq.\ref{gg2}, there is a
  significant difference at large values of $x$ which is around $4-6\%$. 
  
  In Fig.~\ref{fig6}, we present the results for ${\it R^\prime}=\frac{\frac{5}{18} F_{2A}^{Weak}(x,Q^2)}{F_{2A}^{EM}(x,Q^2)}$ 
 vs $Q^2$ at different $x$ for $^{56}Fe$ and compared them with results obtained for isoscalar nucleon. If we assume
 $s(x,Q^2) = \bar s(x,Q^2) = c(x,Q^2) = \bar c(x,Q^2)$ then the ratio ${\it R^\prime}=1$ for isoscalar nucleon. However, in the presence of 
 non-zero PDFs for charm and strange quarks, this ratio ${\it R^\prime}~\ne~1$ in the region of low $x$. This difference is 
 about $2\%$ for the isoscalar nucleon and increases to $6\%$ in the case of $^{56}Fe$ when the full model is used at NLO. This 
 increase in ${\it R^\prime}$ due to nuclear medium and nonisoscalarity effect is almost constant with the increase in $Q^2$.
\section{Conclusion} \label{sec:Summary}
Recent results of a phenomenological analysis of nuclear structure functions by the CTEQ 
collaboration~\cite{talknuint} suggesting that $F_2^{EM}(x,Q^2)$ may be different from
$F_2^{Weak}(x,Q^2)$ in iron nucleus specially at low $x$, have been examined in a microscopic model by studying the nuclear medium effects. The model makes use of
local density approximation to study the effect of Fermi motion, nuclear binding and nucleon correlation effects using a relativistic spectral function
and also includes the contribution of pion and rho mesons. The shadowing effect has been included following the work of Kulagin and Petti~\cite{Kulagin:2004ie}. 
The results have been 
compared with the experimental data. It is concluded from the study that:
\begin{itemize}
 \item The structure functions  $F_{2 A}^{EM}(x,Q^2)$  and  $F_{2 A}^{Weak}(x,Q^2)$ in iron nucleus are almost same for large $x$ i.e $x~>~0.3$,
 but are different in small $x$ region($x<0.3$). This difference is found to be very small for isoscalar nuclei, and is similar to the case 
  of free nucleon, where it is expected to exist due to the presence of heavy flavor quark-parton distribution functions, which are different in 
  $F_2^{EM}(x,Q^2)$  and  $F_2^{Weak}(x,Q^2)$ structure functions. Quantitatively similar results are obtained even in the presence of mesonic and shadowing
 effects.
 
 \item In the region of low $x(x~<~0.3)$, the differences in $F_2^{EM}(x,Q^2)$  and  $F_2^{Weak}(x,Q^2)$ structure functions are mainly due to
 nonisoscalarity of nuclei which is about $\sim 4\%$ in $^{56}Fe$ and increases 
 to $\sim 7\%$ in $^{208}Pb$ at $x=0.1$.
 
 \item The present results are found to be in good agreement with the experimental results from JLab~\cite{Mamyan:2012th}, NuTeV~\cite{Tzanov:2005kr}
 and CCFR~\cite{Oltman:1992pq} but are higher than the older data
  of EMC~\cite{Aubert:1986yn} and CDHSW~\cite{Berge:1989hr} which are at low $Q^2$. The results are in quantitative agreement with the preliminary 
  results of Kalantarians~\cite{talknuint} where a phenomenological analyses of $F_2^{EM}(x,Q^2)$  and  $F_2^{Weak}(x,Q^2)$ are performed.
  It should be noted that in the region of low $Q^2$ other non-perturbative effects in the 
  nuclear medium like the dynamic higher twist correction arising due to quark-quark, quark-gluon interactions and coherent meson 
  interactions in the nucleus may be important~\cite{Yang:1998zb,Miller}. Such effects are beyond the scope of this work and are not considered
   here.
\end{itemize}

The theoretical results presented in this paper show the difference between $F_2^{EM}(x,Q^2)$  and  $F_2^{Weak}(x,Q^2)$ are quite small  
 at large $x(x>0.3)$  and could be around $4-6\%$ in the region of very small $x$(and $Q^2$) in the case of $^{56}Fe$. 
 The available exprimental results on $F_2^{EM}(x,Q^2)$  and  $F_2^{Weak}(x,Q^2)$ in this region of $x$ and $Q^2$ could overlap with each other 
 when statistical and systematic errors are taken in to account. However, presently ongoing DIS experiments at JLab~\cite{jlabupdate}
 using charged lepton and MINERvA~\cite{Tice:2014pgu} using 
neutrino/antineutrino beams with several nuclear targets in the region of low $x$ and $Q^2$, it should be possible to determine more precisely the difference 
 between  $F_2^{EM}(x,Q^2)$  and  $F_2^{Weak}(x,Q^2)$ structure functions. The results presented in this paper will be helpful in understanding the physics of 
 these differences in a microscopic model.

 \begin{acknowledgments}
  M. S. A. is thankful to Department of Science
and Technology(DST), Government of India for providing financial assistance under Grant No. SR/S2/HEP-18/2012. 
I.R.S. thanks MINECO(Spain) for supporting his research under a Juan de la Cierva-incorporacion contract.
 \end{acknowledgments}

\end{document}